\documentclass{aa}  

\usepackage{graphicx,epsfig,natbib}
\usepackage{txfonts}
\begin{document}
   \title{Study of stellar populations in the bulges of barred galaxies\thanks{Based on observations obtained at Siding Spring Observatory (RSAA, ANU, Australia) and the INT telescope at the ING, La palma, Spain}}
   

     \author{I. P\'erez \inst{1,2}\and
          P. S\'anchez-Bl\'azquez\inst{3,4,5,6}}
  
 \institute{Dep. F\'isica Te\'orica y del Cosmos, Campus de Fuentenueva, Universidad de Granada, E-18071 Granada,
 Spain \\  \email{isa@ugr.es} \and Instituto Carlos I de F\'isica Te\'orica y Computaci\'on, Spain \and Dep.
 F\'isica Te\'orica, Universidad Autonoma de Madrid, Cantoblanco, E-28049, Madrid, Spain \and Instituto de
 Astrof\'{\i}sica de Canarias, c/V\'{\i}a L\'actea s/n, E-38200, La Laguna, Tenerife, Spain \and Departamento de
 Astrof\'{\i}sica, Universidad de La Laguna, E-38205 La Laguna, Tenerife, Spain \and Jeremiah Horrocks
 Institute, UCLan, Preston, PR1 2HE, UK \\ \email:{p.sanchezblazquez@uam.es}
               }

   \offprints{I. P\'erez}

   \date{Received September 15, 1996; accepted March 16, 1997}

 
  \abstract
   {}
   {To determine the influence of bars in the building of galaxy bulges through the analysis of ages and metallicities derived from stellar absorption line-strength indices.}
   {Long-slit spectroscopy was obtained  for  a sample 20 early-type barred galaxies. Line
strength indices were measured and used to derive age and metallicity gradients in the bulge region by comparing with  stellar population models.
 The same analysis was carried out with similar data of unbarred galaxies taken from the literature.}
   {The bulges of barred galaxies seem to be more metal rich, at a given velocity dispersion ($\sigma$), than the bulges of unbarred galaxies, as measured by some metallicity sensitive indices. There are indications that the ratio of relative abundance of alpha-elements with respect to iron,  [E/Fe], derived for the bulges of barred galaxies tend to lie above the values of the unbarred galaxies at a given $\sigma$. The metallicity gradients for the majority of  the bulges are negative, less metal rich towards the end of the bulge. The gradient values show a large scatter for galaxies with $\sigma$ below 150 km\,s$^{-1}$\,.  The age distribution is related to the presence of bulge substructure such as a nuclear ring or an inner disk. The metallicity of both the bulge and the bar are very well correlated indicating a close link between the enrichment histories of both components. }
   {Bulges of barred early type galaxies might have suffered a different chemical enrichment compared to the bulges of unbarred galaxies of the same morphological type, same central velocity dispersion and low inclination angles. The hinted stellar populations differences separating the bulges of barred and unbarred galaxies and the strong link existing between the metallicity of the bulge and the presence of a bar points to scenarios were they both form simultaneously in processes leading to rapid and massive episodes of star formation, possibly linked to the bar formation.  In order to confirm and generalise the results found here, it would be useful to extend the data set to a larger number of unbarred galaxies and a wider range of morphological types}


   \maketitle
%

\section{Introduction}

If we consider that galaxy bulges of disk galaxies are the excess of light from the exponential disk, excluding the bar (e.g. Freeman 1970; Fisher 2006; Peletier 2008),\nocite{freeman,fisher06,peletier08} we can find two types of 
bulges. In the standard picture, the two types of bulges are separated between those formed through violent processes 
($'$classical bulges$'$) and those formed slowly through internal processes (i.e., through secular evolution) (e.g. Simkin et al. 1980; Pfenniger \& Friedli 1991; Hopkins et 
al. 2010)\nocite{simkin1980,pfenniger1991,hopkins10} 
They are expected to have very different structural 
properties; the bulges formed via secular evolutionary processes are expected to have more disk--like properties while 
the so called 'classical' bulges are expected to have properties more related to elliptical galaxies (e.g. Fisher 
\& Drory 2008; Gadotti 2009; Fisher \& Drory 2010,\nocite{fisher08,gadotti09,fisher10} for a detail discussion on the 
properties on both types of bulges, see the review by Kormendy \& Kennicutt 2004 and referencies therein).  However, 
both bulges can coexist and the observed properties sometimes depend on the geometry chosen to derive the bulge 
properties.\nocite{KKreview}

Much work has been done trying to characterise the influence of the disk in the bulge evolution
by comparing the stellar populations of bulges and those of elliptical galaxies and, up to date, there is not
consensus in the final conclusions.
Some  authors (e.g. Thomas \& Davies 2006\nocite{thomas06}; Goudfrooij, Gorgas \& Jablonka 1999\nocite{goudfrooij}) have 
concluded that, galaxies with morphological types from E to  as late as Sbc, have stellar population properties 
that are correlated to the central velocity 
dispersion ($\sigma$) and that, at a given $\sigma$, bulges and elliptical galaxies cannot be distinguished as far as 
their stellar populations are concerned, indicating very little influence of the disc in the evolution of the bulge. Recent work by MacArthur, Gonzalez \& Courteau (2009)\nocite{macarthur09} derived the whole star formation histories for the bulges of a sample of eight galaxies, finding that the mass weighted stellar ages of the bulges are old and that, therefore, secular evolution effects only 
contribute minimally to the total bulge mass.  

However, other authors (e.g.Falc\'on-Barroso et al. 2002\nocite{falconbarroso}; Peletier et al. 2007\nocite{peletier07}; Ganda et al. 2007\nocite{ganda}) find that the
Mg--$\sigma$ relation for the bulges lies below that found for elliptical galaxies.  Proctor \& Sansom
2002\nocite{proctor} go further concluding that the stellar population trends greatly differ between early and late
type galaxies.


Part  of the discrepancy between different studies can be due to the selection of the sample. 
For example, inclination effects can make the contribution to the integrated spectra of the different galaxy components appear larger or smaller depending on the inclination (see Peletier et al. 2007). Furthermore, the morphological selection and the 
range in luminosities chosen might bias the conclusions 
as kinematical studies indicate that secularly formed bulges are more important in late type spirals and 
especially in low-luminous ones.

An alternative way to determine the influence of secular evolution in the growing of bulges is to 
compare similar spiral galaxies (in terms of luminosity, morphology and inclination)  with and without bars.
Bars are
expected to play an important role in the secular evolution of disk galaxies and 
they are obvious candidates to produce secular evolution and create a bulge secularly in disk galaxies. 
The bar gravity torques make the gas lose 
angular momentum which, in turn, provokes an inflow of gas towards the central parts  (e.g. Pfenniger \& Norman 
1990; Combes et al. 1990; Friedli \& Benz 1995). Star formation might be triggered and, possibly, the formation of a stellar 
bulge.\nocite{pfenniger1990,combes1990,friedli95} One obvious way to study the efficiency of this process is to compare 
the star formation history of bulges in galaxies with and without bars. These type of studies have to be done in face-on galaxies where it is easy to morphologically identify the bar.  There have been only a handful of studies 
characterising the bulges from samples of face-on galaxies. 39 galaxies were analysed by Moorthy \& Holtzman (2006) 
\nocite{moorthy}. From these, 19 have low inclination, and eleven are barred. They found that bars have smaller 
H${\beta}$ (indicating older luminosity-weighted ages) that unbarred galaxies at a fixed central velocity dispersion 
and maximum rotational velocity. They also analysed the distribution of the stellar populations with radius, finding 
that, when positive age gradients exists, they were always found in barred galaxies. 
On the contrary, studies focused only on edge-on galaxies attempting to classify bars such 
as Jablonka et al. (2007)\nocite{jablonka2007} do not find any difference between the indices in barred and unbarred 
galaxies, neither in the central values nor in the gradients. 

We have started a project to characterise the stellar properties of barred galaxies to understand the influence of bars 
in the evolution of the different galaxy components and to study their formation. In P\'erez, S\'anchez-Bl\'azquez \& 
Zurita (2007)\nocite{perez07} and P\'erez, S\'anchez-Bl\'azquez \& Zurita (2009, hereafter, Paper I), we presented a 
detailed analysis of the stellar population parameters along the bar of a sample of 20 galaxies. From the results 
obtained (kinematics and SSP derived stellar population parameters along the bar) we determined that the bulge 
kinematics is closely linked to the presence of a bar. Some studies, however, have not found any correlation between 
the presence of a bar and the bulge properties, probably due to an inclination bias because studies of more inclined 
objects will avoid the contribution of the inner disc--like components. To conclude whether bars have a strong 
influence in the properties of bulges, a systematic comparison of bulges of barred and unbarred galaxies needs to be 
done, using galaxies with similar inclination, morphological type and central velocity dispersion.  We present, 
in this paper, the study of the stellar population properties of the central regions of the 20 barred galaxies analysed in Paper 
I. For an analysis of the differences between the bulges of barred and unbarred galaxies our sample would need to be complemented with a similar sample of galaxies without bars. So, in addition to the bulges of barred galaxies,  we present, in this paper, a comparison with Moorthy et al. (2006) data from a sample of both, barred and unbarred galaxies, being the only sample available in the literature with similar characteristics to our data; namely, spectral coverage and resolution, morphological type and inclination values. The data have been re--analysed in the same way as for our sample. This is the first time that such study is done for a large sample of barred galaxies, with a total of 31 barred galaxies combining both samples. Due to the low number of unbarred galaxies compared to the barred galaxies in the total sample, we will extend the sample in the future to include a larger number of unbarred galaxies and to cover a wider range in morphological types.
 
In Secs.~\ref{sample}, \ref{observations} and ~\ref{line} we give a brief introduction to the sample selection, 
observations and, data reduction and line-strength measurements. A more detailed explanation of these sections is given 
in Paper I. An analysis of the central line--strength indices and a comparison between barred and unbarred galaxies is 
given in Sect.~\ref{central.line.strengths}. The line--strength index distributions in the bulge region are presented 
in Sect.~\ref{line.strength.distribution.gradients}. The single stellar population equivalent parameters are discussed 
in Section~\ref{age-metallicity}. The age and metallicity gradients are presented in 
Section~\ref{age.and.metallicity.gradients}, while a comparison between the stellar parameters in the bulge and bar 
region is given in Sect.~\ref{comparison.bar.bulge}. The conclusions and discussion are presented in 
Sect.~\ref{discussion}.

 
 \section{Sample characterisation}
 \label{sample}
 
We have selected barred galaxies from the Third Reference Catalogue of bright galaxies (RC3)
 (de Vaucouleurs 1948)\nocite{vaucouleurs} with  the following criteria; to be classified as barred, with inclinations between 10$^{\circ}$ and 70$^{\circ}$, and nearby ($cz$~$\le$~4000~kms$^{-1}$).
 The sample is biased towards early-type barred galaxies, which have higher surface brightness. This morphological criteria was set to ensure that we could obtain data with enough signal-to-noise in the bar region (see Paper I)\nocite{psz}. In the future, we plan to enlarge the sample towards later types to analyse the trends with morphological type. Half of our galaxies have nuclear activity to analyse possible trends between the nuclear activity, the bar characteristic of the bar and the bulge. Furthermore, eight of the galaxies present nuclear bars. Our final sample comprises 20 galaxies. For commodity, we reproduce here the Table (Table~\ref{sample.tab}), already shown in Paper I, with the main characteristics of the sample as taken from the Hyperleda catalogue\nocite{paturel} (Paturel et al.\ 2003)\footnote{http://leda.univ-lyon1.fr}. The bar strength shown in Table~\ref{sample.tab}\footnote{ Bar strength is shown here as bar--class; for assignment of bar--class to a certain bar strength, see Buta \& Block 2001, it is on a scale from 0 to 6, 6 being the strongest bar\nocite{buta2001}} has been taken from the literature, where strength is defined as the torque of a bar embedded in its disk (Combes \& Sanders 1981\nocite{combes81}), see Table~\ref{sample.tab} for the references for the individual galaxies. The nuclear types have been obtained from~\cite{veron}. The sample shows a wide distribution of maximum rotational velocities (80 - 260 kms$^{-1}$).

 As a working definition, we use a bulge size based on the kinematic and the line profile information.  We consider the end of the bulge region the radius at which the $\sigma$ starts decreasing after a plateau or dip. In Paper I, we noticed that it also coincides with a change in the slope of the line-strength profiles for all the indices. After analysing broad band images, as in Paper I, and also the HST images published in the literature (e.g. Comer\'on et al. 2010)\nocite{comeron}, we have also seen that this region coincides with changes in the morphology. Therefore, the bulge end defined using the kinematics is almost always coincident with the end of the nuclear structure; nuclear ring, double bar, inner spiral, etc; found in the broad-band images. Table~\ref{parameters} shows the approximate bulge and bar sizes, the latter given by the minimum ellipticity in the bar region (see Paper I for a further explanation). In Table~\ref{parameters} we give the central structure and bar semi-major axis size values only for those galaxies for which we have good broad-band imaging, see Paper I for a more detailed explanation. In the same paper, we present the line-of-sight position diagrams and velocity dispersion for the galaxies (figure 8). From a close examination of the stellar kinematics in the bulge region we showed that all the galaxies in the sample have disk-like structures in their centres.

 Due to the presence of these disk-like structures in our galaxies, we have used, in this work, a maximum velocity dispersion instead of a central velocity dispersion. This velocity dispersion is defined as the maximum velocity dispersion in the bulge region. We have chosen this dispersion value to avoid being dominated in some galaxies by the low central velocity dispersion that could be indicative of nuclear star formation (e.g nuclear disk). Therefore, the maximum velocity dispersion should be closer to the velocity dispersion of the bulge.

  \section{Observations and reduction}
  \label{observations}
  
We obtained long-slit spectra along the bar major-axis for our sample of 20 barred galaxies. The bar position angles 
were derived using the Digital Sky Survey (DSS) images. The observations were performed in two different runs, with the Double Beam Spectrograph  at Siding Spring Observatory (Australia) and with the IDS spectrograph at the Isaac Newton Telescope (La Palma, Spain), respectively. 
The observations were described in detail in Paper I. To summarise,
in the first run, spectra for 6 galaxies were obtained, covering a wavelength range from 3892-5815\AA\ and a spectral resolution of FHWM$\sim$2.2\AA. In the second run, spectra for 14 galaxies were obtained, covering a wavelength range of 3020-6665\AA\ and a spectral resolution of $\sim$3\AA~(FWHM).

The reduction of the two runs was carried out with the package REDUCEME (Cardiel 1999)\nocite{cardiel}.  Standard data reduction procedures (flat-fielding, cosmic ray removal, wavelength calibration, sky subtraction and fluxing) were performed. Error images were created at the beginning of the reduction and were processed in parallel with the science images. For details about the reduction steps see Paper I.

In order to derive the index spatial distribution for the fully reduced galaxy frames, a final frame was 
created by extracting spectra along the slit, binning in the spatial direction to guarantee a minimum 
signal-to-noise ratio of 20 per \AA~ in the spectral region of Mgb. This minimum S/N ensures errors lower 
than 15\% in most of the Lick/IDS indices (Cardiel et~al.\ 1998)\nocite{cardiel98}.
A careful emission-line removal was performed with {\tt GANDALF} (Sarzi et al. 2006)\nocite{sarzi}. The details of the procedure can be found in Paper~I. This emission-line removal procedure fits simultaneously, the stellar and the emission line spectra, by treating the  emission lines as additional Gaussian templates and iteratively searching for the best radial velocity and velocity dispersion. When the nebular emission-line profiles are clearly asymmetric  (i.e in the nuclei of active galaxies) and cannot be fitted by a single Gaussian, double Gaussians are use to fit the emission lines. 

\begin{table*}
\begin{center}
\caption{General properties of the sample\label{sample.tab}}
\begin{tabular}{l r l c l l r r r }     
\hline\hline
Object          & v(km~s$^{-1}$)    & Type & Bar-class$^{1}$& Nuclear type & Inner morph.$^{2}$&$B$& V$_{max,gas}$  (km\,s$^{-1}$\ )$^{3}$& $i$(deg)\\
\hline
NGC~1169$^{a}$   &2387& SABb    &3          & --             &   --         &12.35        & 259.1 $\pm$ 7.3              & 57.1 \\
NGC~1358            &4028& SAB(R)0    &    --       &Sy2           &   --         &13.19        & 136.1 $\pm$ 10.6              & 62.8 \\
NGC~1433$^{b}$   &1075&(R)SB(rs)ab&4&Sy2&Double--bar$^{a}$&10.81& 85.1 $\pm$ 2.4 & 68.1\\
NGC~1530$^{b} $  &2461&SBb&6&--&--&12.50&169.1$\pm$3.5&58.3\\
NGC~1832$^{d}$   &1939&SB(r)bc&2&--&--&12.50&129.9$\pm$2.0& 71.8\\
NGC~2217            &1619 &(R)SB(rs)0/a&--&LINER?&Double--bar$^{b}$&11.36& 183.4$\pm$ 9.2 & 30.7\\
NGC~2273$^{c}$   &1840&SB(r)a&2&Sy2&--&12.62&192.2$\pm$5.5&57.3\\
NGC~2523            &3471&SBbc&--&--&--&12.64 &211.4 $\pm$10.9 &61.3\\
NGC~2665            &1734&(R)SB(r)a&--&--&--&12.47& 130.9$\pm$7.1&32.8\\
NGC~2681$^{c}$   &692  &(R)SAB(rs)0/a&1&Sy3&Triple--bar$^{c}$&11.15&87.5$\pm$6.7&15.9\\
NGC~2859$^{c}$   &1687&(R)SB(r)0~\^&1&Sy&Double--bar$^{d}$&11.86& 238.5$\pm$13.3& 33.0\\
NGC~2935            &2271 & (R)SAB(s)b  & --& -- &-- &12.26 &188.3$\pm$2.0 &42.7\\
NGC~2950            &1337 &(R)SB(r)0\^~0&--&--&Double--bar$^{e}$&11.93& --& 62.0\\
NGC~2962            &1966 &(R)SAB(rs)0&--&--&Double--bar$^{c}$&12.91&202.9$\pm$9.9& 72.7\\
NGC~3081$^{c}$  &2391  &(R)SAB(r)0/a&3&Sy2&Double--bar$^{d}$&12.89 &99.9$\pm$4.0&60.1\\
NGC~4245$^{c}$  &815   &SB(r)0/a&2&&--&12.33&113.5$\pm$5.4&56.1\\
NGC~4314$^{a}$  &963   &SB(rs)a&3&LINER&Double--bar$^{c}$&11.42&253.3$\pm$24.6&16.2\\
NGC~4394$^{d}$  &922   &(R)SB(r)b&3&LINER&--&11.59& 212.5$\pm$16.0&20.0\\
NGC~4643$^{c}$  &1335 &SB(rs)0/a&3&LINER&--&11.68&171.4$\pm$7.2&42.9\\
NGC~5101$^{d}$  &1868 &(R)SB(r)0/a&2&--&--&11.59&195.7$\pm$9.0&23.2\\
\hline
\end{tabular}
\end{center}
{\footnotesize 
$(1):$
$(a)$ Bar class derived from the $K-band$ light distribution, ~\cite{block}\\
$(b)$ Bar class derived from the $K-band$ light distribution,~\cite{block04}\\
$(c)$ Bar class derived from the $K-band$ light distribution,~\cite{buta} \\
$(d)$ Bar class derived from the $H-band$ light distribution,~\cite{laurikainen}\\
$(2):$
$(a)$~\cite{buta86}
$(b)$~\cite{jungwiert}
$(c)$~\cite{erwin04}
$(d)$~\cite{wozniak95}\\
$(3)$ Rotational velocity corrected for inclination}
\end{table*}

\section{Line-strength indices}
\label{line}

 Lick/IDS line-strength indices using the definition in Trager et al. (1998)\nocite{trager} were measured in 
all the binned spectra, cleaned of emission, along the radius. The errors were calculated from the uncertainties caused by photon noise, wavelength calibration and flux calibration. 

 Line-strength indices depend on the broadening of the lines caused by instrumental resolution and by the internal motion of the stars. Before measuring the indices, we broadened our spectra to the Lick/IDS
wavelength dependent resolution following the prescriptions of Worthey \& Ottaviani (1997)\nocite{worthey97}. Afterwards, we applied a correction due to the velocity dispersion of the galaxies using, for each spectrum, the template obtained in the derivation of the velocity dispersion, as described in Paper~I. 
Offsets to transform into the Lick/IDS spectrophotometric system using stars in common with this library were derived (see Paper~I). However, we only 
apply the offset to the indices when comparing with the Thomas et al. (2003) models based on the Lick/IDS fitting functions.
To study the central and global properties of the bulges, we have also extracted the spectra in 3 different apertures  by averaging line-strength indices inside 1.2, 3.6 and 10 arcsec and a fourth one 
averaging the indices inside the whole bulge, using the sizes given in Table~\ref{parameters}. 
The process is different from  adding the spectra inside those apertures and measure the indices afterwards. We decided to proceed in this way, 
first, to be able to compare with other authors when only line-strength indices were available but not the whole image and, secondly, to magnify the differences
in the external parts of the bulges, which normally do not have much weight when a normal  extraction, necessarily weighted with the light, is performed. The error is calculated as the standard error of the mean which is the standard deviation of the sampling distribution of the mean.

\begin{table}
\centering
\caption{Bar and bulge sizes}
\label{parameters}
\begin{tabular}{l r r}     
\hline \hline

 Object & Bar semi-major axis  &  Bulge radius \\ 
            & (arcsec) & (arcsec) \\
NGC1169&29&  5 \\
NGC1358& -- & 5\\  
NGC1433  & --& 5\\
NGC1530 &69 & 15 \\
NGC1832&-- & 7\\
NGC2217 &-- &  5\\
NGC2523 &-- &  10\\
NGC2665 & --&  5\\
NGC2681& 23&  8\\
NGC2273 & 21&  5\\
NGC2859 & 48&   7\\
NGC2935& -- & 7 \\
NGC2950 & 44&   10\\
NGC2962 & 45&  5\\
NGC3081 &35 &5 \\
NGC4245 &59 &  6\\
NGC4314 & 92&  8\\
NGC4394 &56& 5\\
NGC4643 &67&  10\\
NGC5101& --&7 \\
 \hline
  
\end{tabular}
\end{table}



  
 
\section{Results} 
\label{results}
 In order to analyse the influence of bars in building bulges, in this section we compare the 
line-strength indices in the bulge of barred and unbarred galaxies.
We also analyse the spatial distribution of the indices, as differences may be expected  due to the differences in the gas dissipation 
processes and star formation events in bulges built via mergers and via secular evolution.
We also compare our line-strength indices to stellar population models to derive  SSP-equivalent ages and metallicities both, in central 
apertures and along the radius.

\subsection{Comparison of the central line-strengths of bulges of barred and unbarred galaxies \label{central.line.strengths}}

We have mainly compared throughout the paper our results with those obtained by Moorthy et al. (2006) since the galaxy morphologies are similar in both samples. We selected those galaxies, with b/a between 0.70 and 1.00, and early type galaxies with inclinations from the literature between 10$^{\circ}$ and 70$^{\circ}$, for which no b/a was given in Morthy et al. (2006).  NGC~2787 and NGC~3945 inclinations from Erwin and Sparke (2003)\nocite{erwin2003} and the inclination of NGC~3384 from Busarello et al. (1996)\nocite{busarello}. This selection matches our inclination criteria. Their low-inclination sample contains 19 galaxies, eleven of which are barred. Using these data we can also test, not only the consistency of the results, but also we can search for differences between the barred and the un--barred sample.  B. Moorthy  kindly provided us all the index values along the radius for their sample. This made possible to extract the indices measured in exactly the same way; same resolution (they broadened
their spectra to the Lick/IDS resolution as ourselves), and same apertures as in our own data. We subtract  the offsets they applied to transform their indices to Lick/IDS system. Because their study, as ours, measured the indices in flux-calibrated spectra at the Lick/IDS resolution, 
there should not be additional offsets between their and our line-strengths values. 

\begin{figure*}
 \begin{center}
\resizebox{0.7\textwidth}{!}{\includegraphics[angle=-0]{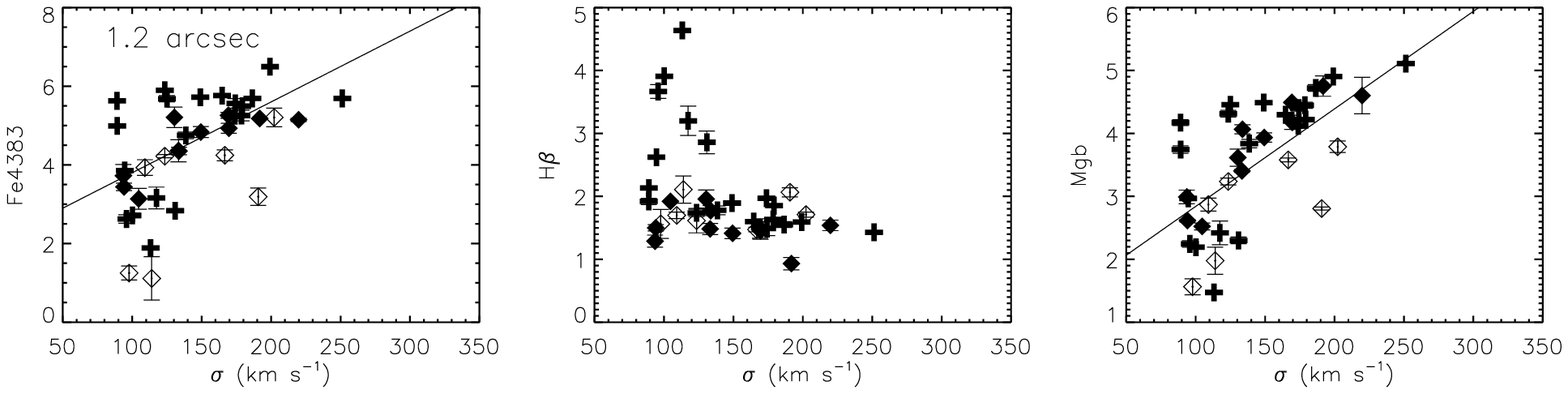}}

\resizebox{0.7\textwidth}{!}{\includegraphics[angle=-0]{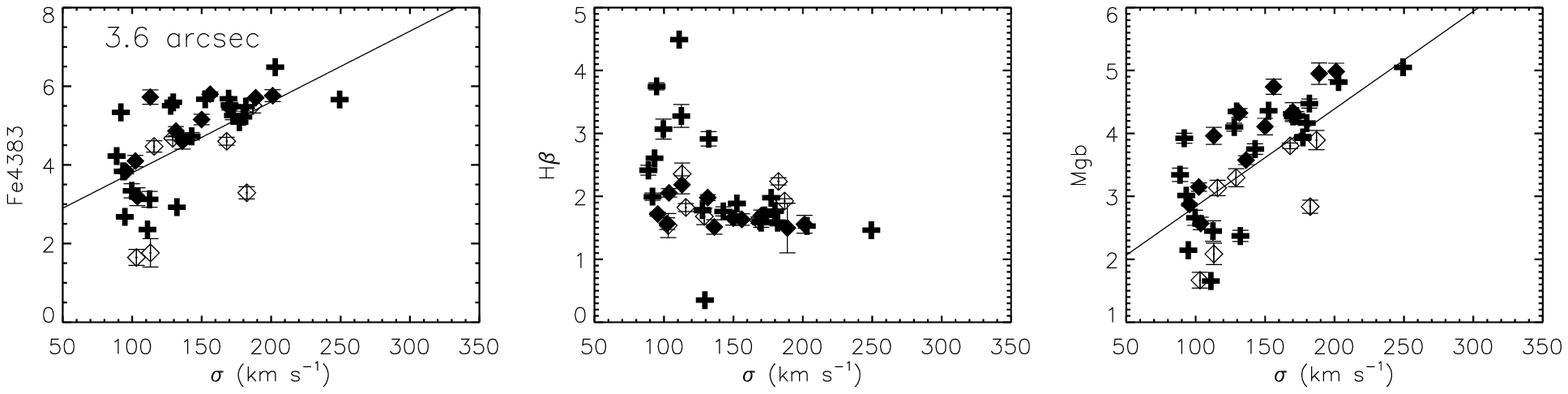}}

\resizebox{0.7\textwidth}{!}{\includegraphics[angle=-0]{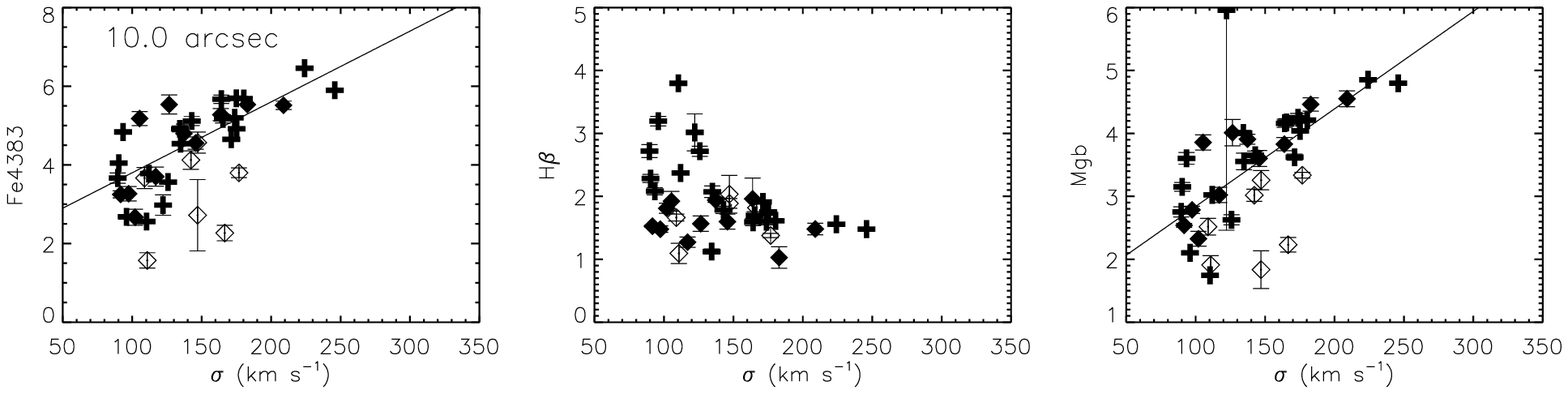}}
        \caption{\label{sigma-1.2}  Distribution of three representative central averaged indices vs. $\sigma$  for our galaxies and Moorthy et al galaxies; 1.2 arcsec average (top panel),  averaged 3.6 arcsec  (middle panel) and averaged 10.0 arcsec   (bottom panel). Empty symbols represent unbarred galaxies, while filled symbols represent barred galaxies with the crosses being our galaxies and the diamonds Moorthy's sample.  }
        \end{center}
\end{figure*}

 Figure~\ref{sigma-1.2} shows the variation of three representative indices inside an aperture of 1.2, 3.6 and 10.0 arcsec vs.  $\sigma$. As it can be seen all  bulges with  
$\sigma>$ 150 km\,s$^{-1}$\  have almost the same index values (except for  Mgb and CN$_2$, not shown in the figure), while 
there is a large scatter in the index-values for bulges with  with $\sigma<$ 150 km\,s$^{-1}$\ 
 This behaviour has already been noticed by several authors in elliptical galaxies  (see, e.g, Caldwell, Rose \& Concannon  2003; Nelan et al. 2005; S\'anchez-Bl\'azquez et al. 2006; Ogando et al. 2006)\nocite{ogando}\nocite{caldwell,sanchez-blazquez2006a,nelan}.  There is a tendency in the metallicity sensitive indices for the barred galaxies to be larger than those derived for unbarred galaxies while the opposite is observed for the Balmer indices.  We have performed a significance test (3 $\sigma$)  on the index distributions with $\sigma$. For some of the indices, namely, the Balmer indices, Fe4383 and Mgb there is a statistically significant difference between the bulges of barred and unbarred galaxies at all apertures while for the other indices the difference is not statistically significant.   

 In Sec~\ref{age-metallicity}  we compare the line-strength with stellar population models to transform the differences in differences in stellar population parameters (name it, age, metallicity ($[$Z/H$]$) and relative abundance of $\alpha$-elements
($[$E/Fe$]$). 

We have not seen that the AGN presence affects the calculated line-strength values after comparing the values for galaxies with and without an AGN nucleus.

    
  
   
     
       \subsection{Comparison of the central line-strengths of bulges and elliptical galaxies}

\begin{figure}
 \begin{center}
\hspace{-1cm}\includegraphics[scale=0.6]{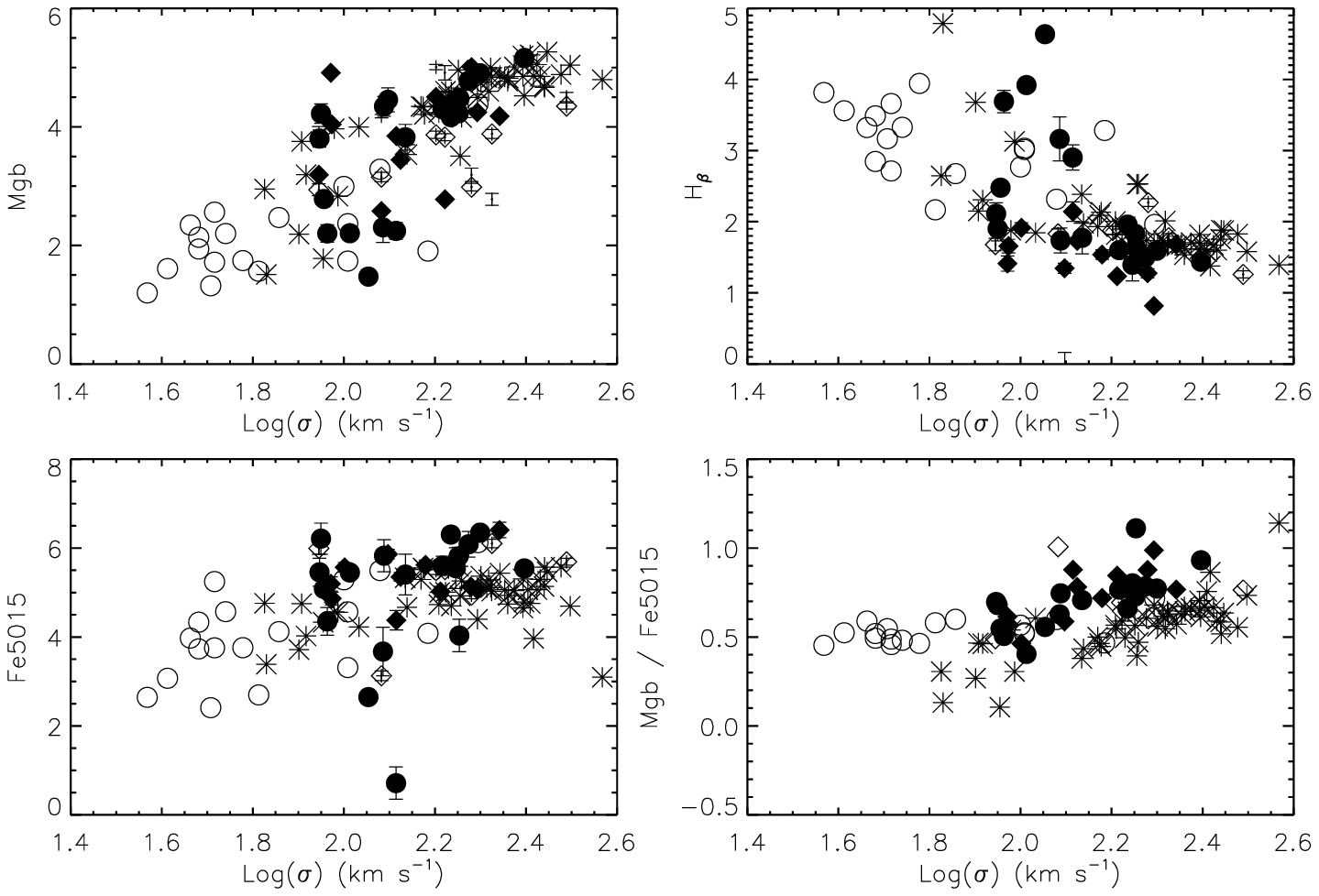}
       \caption{\label{sigma-mg}  Central 1.2 arcsec indices for a reference sample of late spirals, Ganda et al.  (2007), open circles; a sample of ellipticals from Kuntschner et~al.\ (2006), stars; unbarred galaxies used in this work from Moorthy et al.  (2006), open diamonds; barred galaxies used in this work from Moorthy et al. (2006), filled diamonds, and the filled circles represent the values from our sample of barred galaxies. }
    \end{center}
   \end{figure}

Due to the proposed continuation of properties between bulges and elliptical galaxies (e.g. Kuntschner et~al.\  2006), we would like to compare our results with those obtained for elliptical galaxies. Figure.~\ref{sigma-mg} shows the central Mgb, H$\beta$, Fe5015 and Mgb/F5015 values against  the $\sigma$, together with the values presented in Kuntschner et~al.\ (2006)\nocite{kuntschner06} for a sample of ellipticals and 
S0's. For reference, we have also plotted the values for the late-type galaxies in Ganda et al. (2007), which covers the same indices and spectral resolution as  Kuntschner et~al.\ (2006) data. We have also plotted the sample by Moorthy et al. (2006) studied here. There  is a tight correlation between $\sigma$ and Mgb. The trend of H$\beta$, Fe5015 indices with $\sigma$ for all the galaxies is almost flat, showing a large dispersion for galaxies with maximum velocity dispersion below 150km\,s$^{-1}$\,.  The spread is also larger for the galaxies classified as barred. The central Fe5015 index of the unbarred galaxies seems to follow more closely the values of ellipticals while the Mgb follows the opposite trend, with barred galaxies showing values more similar to those of elliptical galaxies. This results in a Mgb~/~Fe5015 ratio where the central values of unbarred galaxies follow more closely those of ellipticals while the barred galaxies tend to lie above those values for the larger central velocity dispersion galaxies.  Those galaxies showing the lower Mgb values are also the ones with lower Balmer indices. 

It is interesting to notice that although Thomas \& Davies (2006)\nocite{thomas06} found that, at fixed  $\sigma$, ellipticals and the bulges of early--type galaxies are indistinguishable, their sample did not contain any barred galaxies. However, Moorthy et al. (2006) found that for a  given $\sigma$ the bulges of barred and unbarred galaxies show differences similar to those found by us with a larger sample.  The implications of these results will be discussed in Section~\ref{discussion}.

\subsection{Line-strength distribution gradients}
\label{line.strength.distribution.gradients}

The variation of the indices with radius in the bulge region was presented in Paper I. The distribution of the indices in the bulge regions is, in some of the cases, far from a linear one. The distribution of the Balmer index is generally characterised by a maximum in the bulge region, close to the bulge end. These profiles are due to the presence of central structures such as nuclear rings. Despite this non--linear distribution and in order to quantitatively characterise the gradients, we have performed a linear fit to the data in the bulge region. Tables in Appendix~\ref{line-strength} show the gradients resulting from the fit to the indices for each of the galaxies.  Most galaxies show negative or no gradients in their metallicity sensitive indices and positive or null gradients in their Balmer indices. NGC~2273 and NGC~2665 show a small positive gradient in their metallicity sensitive indices. NGC~1832 and NGC~2217 also show this trend in their metallicity sensitive indices but it is likely due to the presence of an nuclear ring in the outer part of the bulge. In the case of a long--lasting ring we would expect an enrichment of the ring area due to a continuous star formation activity.

Previous studies of line--strength gradients in bulges (Moorthy et al. 2006; Jablonka et al. 2007; Morelli et al. 2008) also find that most galaxies show negative gradients in the metallicity sensitive indices. 
With respect to the age sensitive gradients (i.e. Balmer indices), Moorthy et al. find that in the cases that a positive age gradient is present it is always in a barred galaxy. In our case, where only barred galaxies are analised, we find galaxies with a variety of index distribution with negative, positive or null Balmer indices gradients. The galaxies which show a negative gradient seems to be associated to the presence of a nuclear ring.  

\section{Ages and metallicities\label{age-metallicity}}
 
\begin{figure*}
 \begin{center}
\hspace{-1cm}\includegraphics[scale=0.8]{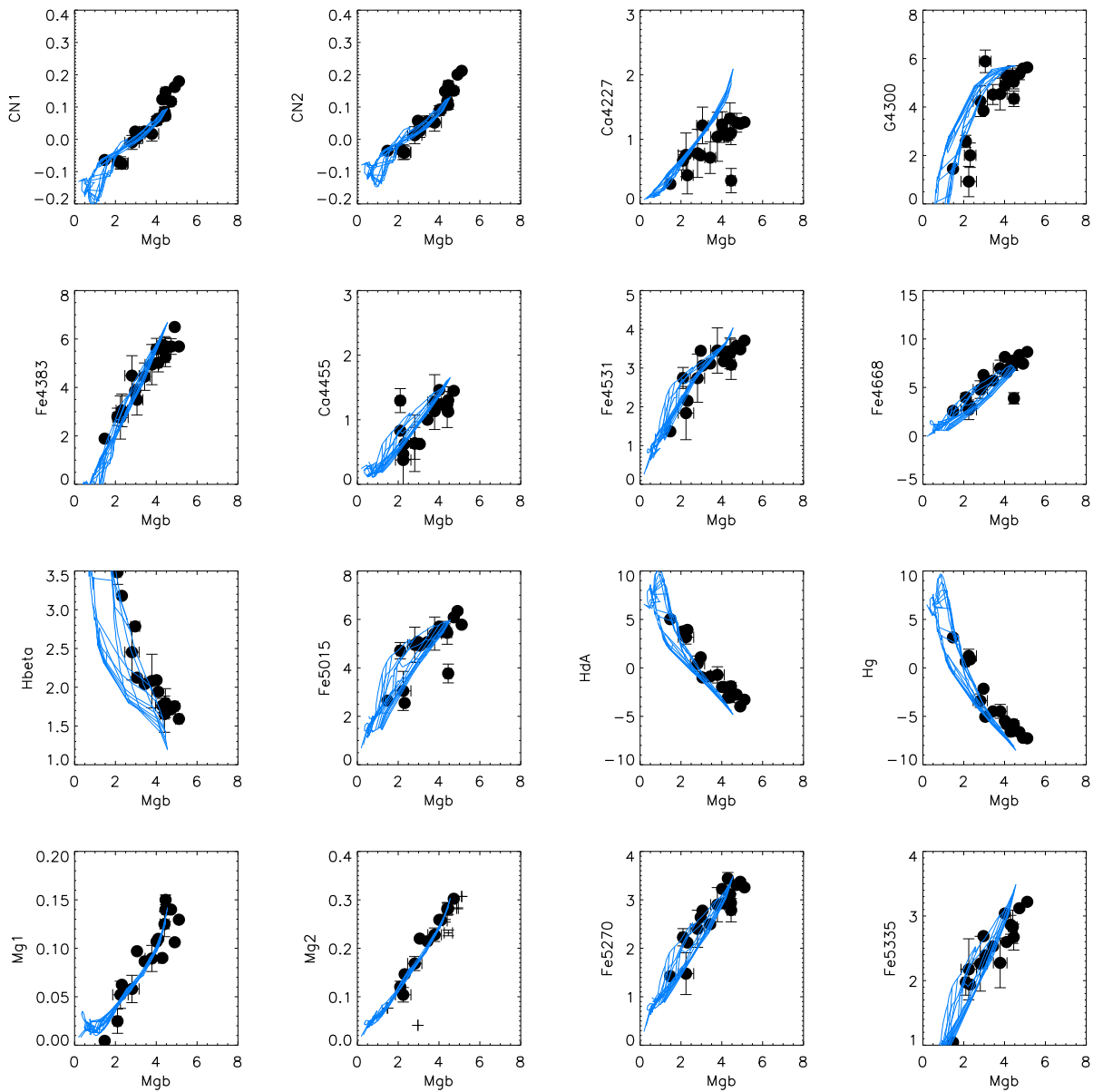}
       \caption{\label{index-mg} Central index vs. central Mgb From top to bottom and from left to right; CN1, CN2, Ca4227, G4300, Fe4383, Ca4455, Fe4531,
C4668, H${\beta} $, Fe5015,H$_{\delta}$, H$_{\gamma}$, Mg1, Mg2, Fe5270, Fe5335. Over-plotted are the models by Vazdekis et al. (2010) for a range of ages (0.5 to 18 Gyr) and 
metallicities (-1.68 to 0.0)  }
      \end{center}
   \end{figure*}

In order to transform our measured indices into single stellar population equivalent parameters, we
use  two different set of models. The first one, from Vazdekis et al. (2010)\footnote{The models are publicly available at http://miles.iac.es}\nocite{vazdekis10}, is built using the MILES library (Cenarro et al. 2007; S\'anchez-Bl\'azquez et al. 2006). \nocite{sanchez06,cenarro07}
The second set of models we use is that of Thomas, Maraston \& Bender (2003)\nocite{thomas2003} which are based on the Lick/IDS fitting functions (Gorgas et al.\ 1993; Worthey 1994)\nocite{worthey,gorgas}. These two set of models use different 
isochrones and stellar libraries and the comparison between the results obtained with them can be used as an estimation of the errors affecting the predictions. The details about the uncertainties are carefully discussed in Paper I. It is clear that the true star formation history in these objects will be more complicated than a simple SSP. However, very useful information can still be retrieved from the SSP analysis if they are properly interpreted. Roughly, the SSP-equivalent age of a composite stellar population (CSP) is dominated primarily by the age of the youngest component and the mass fraction of the different populations. The SSP-equivalent metallicity is very similar to a $V-band$  luminosity weighted chemical composition (Serra \&Trager 2007)\nocite{serra}. 
Fig.~\ref{index-mg} shows the central line indices vs. the central Mgb,  with the Vazdekis et al. (2010) models overplotted. The plot shows a tight correlation among the different 
indices for the central region and very good agreement with the models except for some well-known exceptions.  In particular, some galaxies
show values of CN, Mgb, G4300 and C4668 that are larger than the models while this is not true for the Fe-sensitive indices.  This can be reflecting and overabundance of Mg/Fe and probably C/Fe and N/Fe
with respect to the solar neighbourhood, as suggested by many authors (e.g. Worthey 1998; Tantalo et al 1998; Trager et al. 2000; S\'anchez-Bl\'azquez et al. 2003; Graves \& Schiavon 2008)\nocite{worthey1998,tantalo1998,trager2000,graves2008,sanchezblazquez2003}. Similarly to Paper I, to obtain stellar population parameters we use 4 indices, namely; H${\delta}$A, H${\gamma}$A, Fe4383 and Mgb and follow a multi-index approach as described in Proctor \& Samson (2002), a detailed explanation of the error calculation is given in Paper I. 

\begin{figure*}
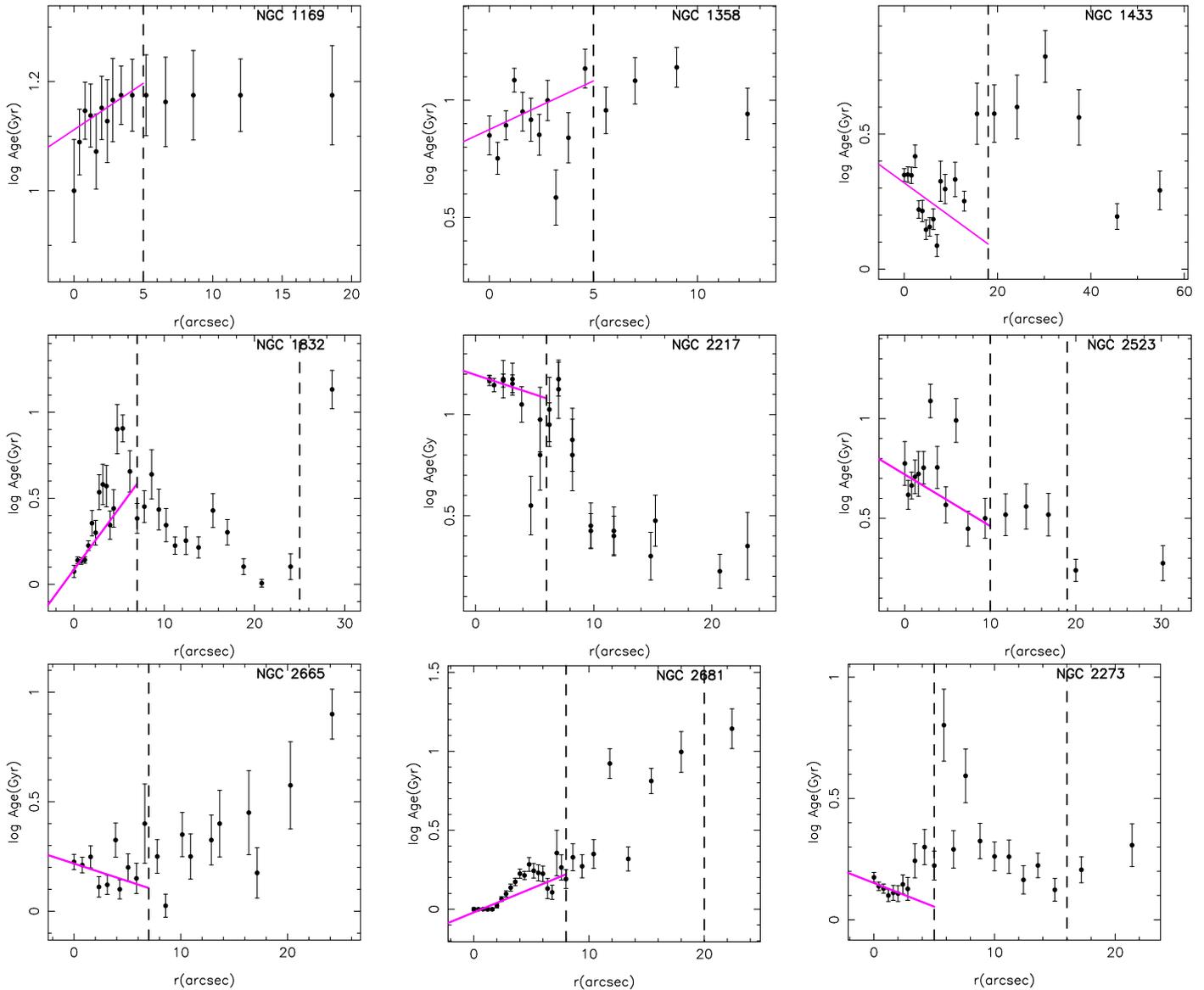

\resizebox{0.3\textwidth}{!}{\includegraphics[angle=-90]{n1169.age.bulge.ps}}
\resizebox{0.3\textwidth}{!}{\includegraphics[angle=-90]{n1358.age.bulge.ps}}
\resizebox{0.3\textwidth}{!}{\includegraphics[angle=-90]{n1433.age.bulge.ps}}
\resizebox{0.3\textwidth}{!}{\includegraphics[angle=-90]{n1832.age.bulge.ps}}
\resizebox{0.3\textwidth}{!}{\includegraphics[angle=-90]{n2217.age.bulge.ps}}
\resizebox{0.3\textwidth}{!}{\includegraphics[angle=-90]{n2523.age.bulge.ps}}
\resizebox{0.3\textwidth}{!}{\includegraphics[angle=-90]{n2665.age.bulge.ps}}\hspace{0.6cm}
\resizebox{0.3\textwidth}{!}{\includegraphics[angle=-90]{n2681.age.bulge.ps}}\hspace{0.6cm}
\resizebox{0.3\textwidth}{!}{\includegraphics[angle=-90]{n2273.age.bulge.ps}}
 \caption{SSP-equivalent age and metallicity along the radius. Dashed lines indicate 
 the end of the bulge and the bar region respectively. A linear fit to the 
 bulge region is also plotted. \label{fig.age.grad}}
\end{figure*}
\addtocounter{figure}{-1}
\begin{figure*}
\resizebox{0.3\textwidth}{!}{\includegraphics[angle=-90]{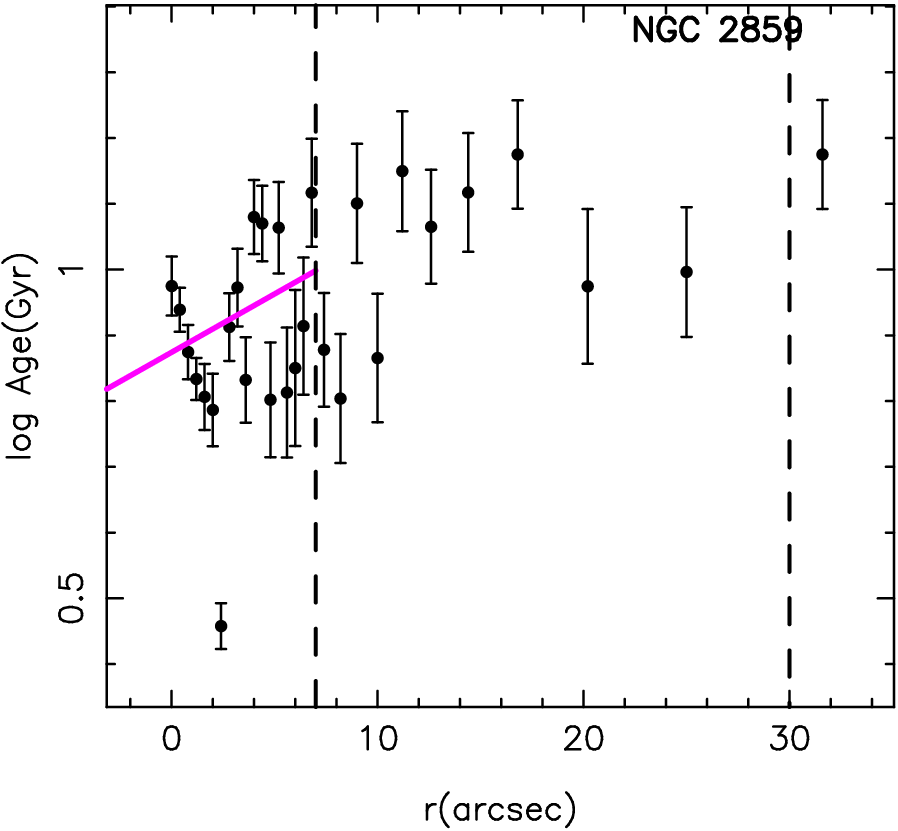}}
\resizebox{0.3\textwidth}{!}{\includegraphics[angle=-90]{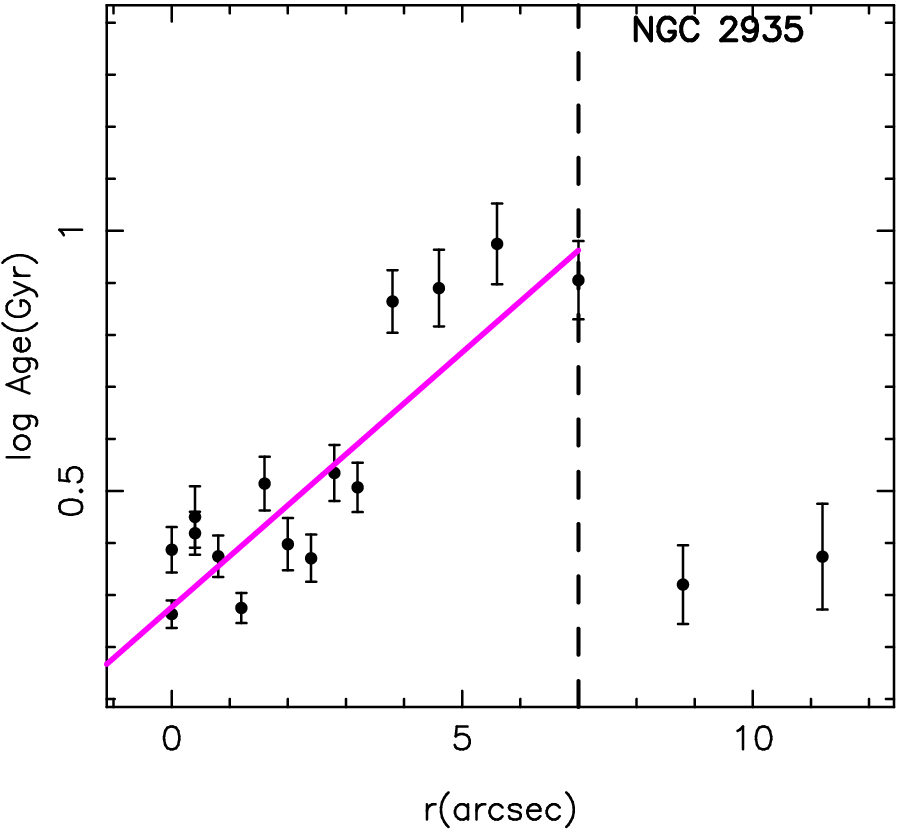}}
\resizebox{0.3\textwidth}{!}{\includegraphics[angle=-90]{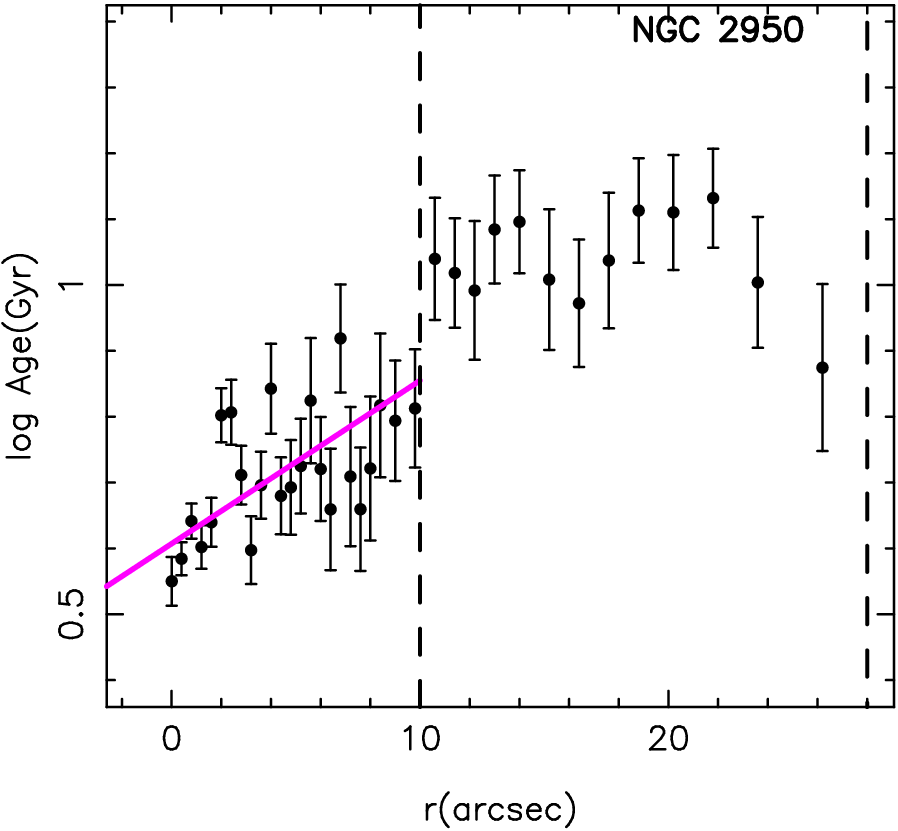}}
\resizebox{0.3\textwidth}{!}{\includegraphics[angle=-90]{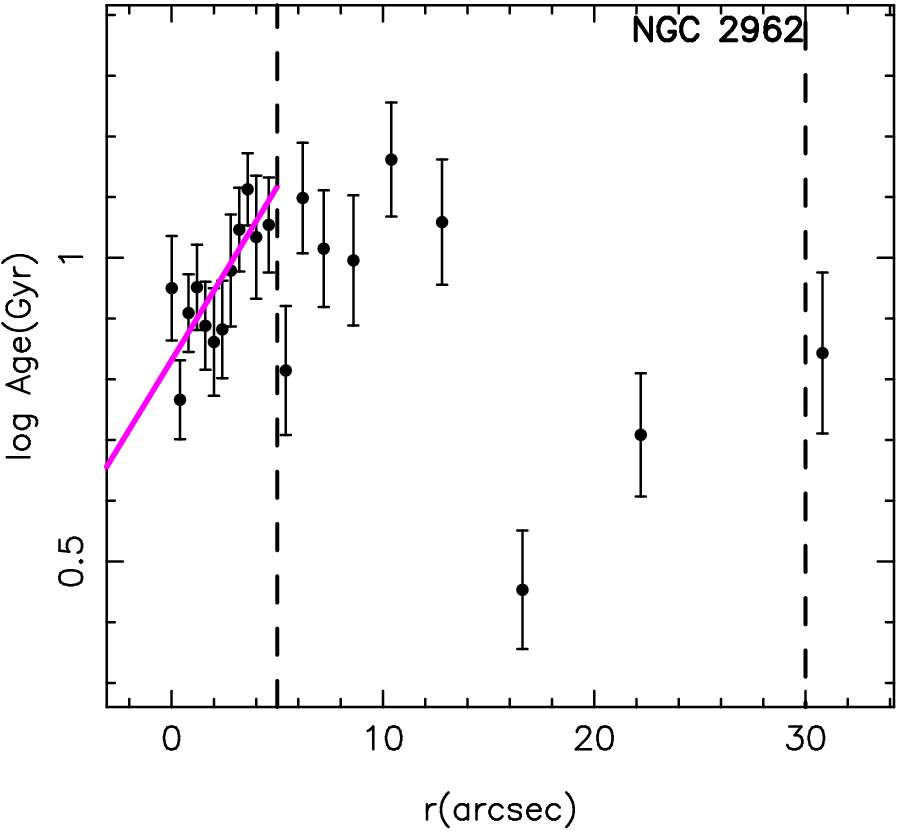}}
\resizebox{0.3\textwidth}{!}{\includegraphics[angle=-90]{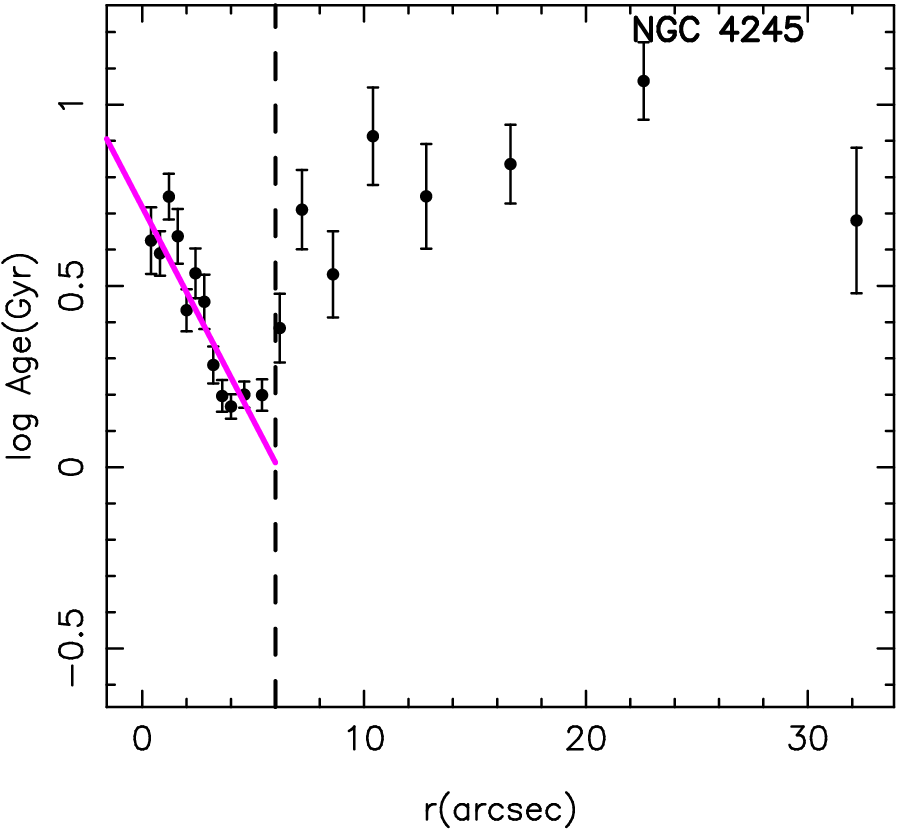}}
\resizebox{0.3\textwidth}{!}{\includegraphics[angle=-90]{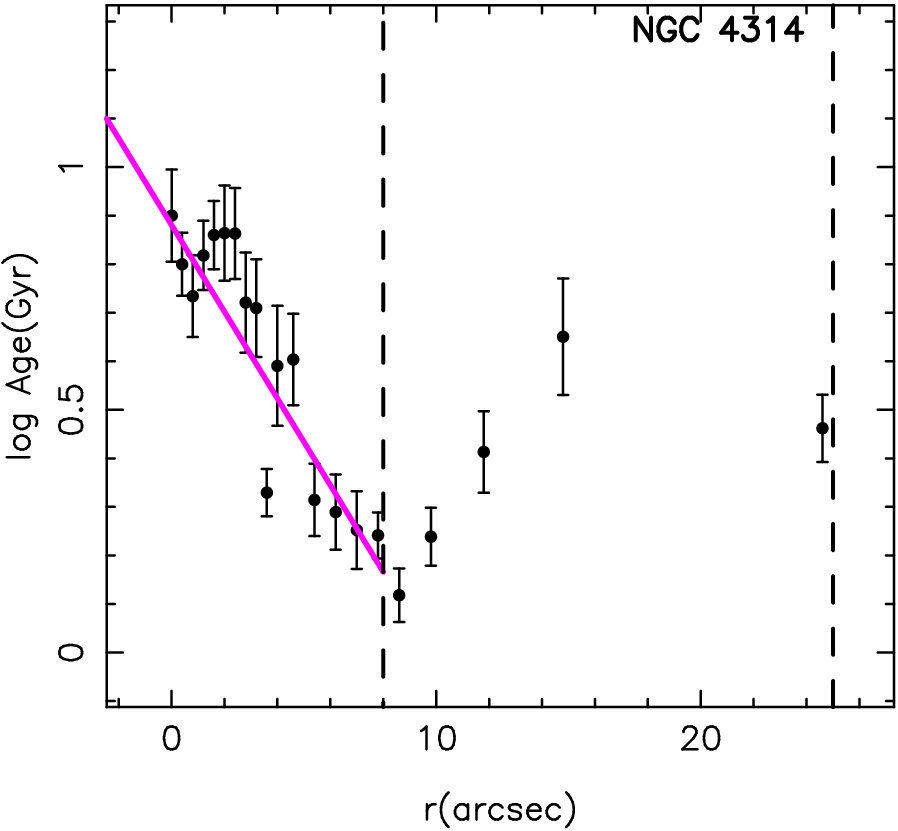}}
\resizebox{0.3\textwidth}{!}{\includegraphics[angle=-90]{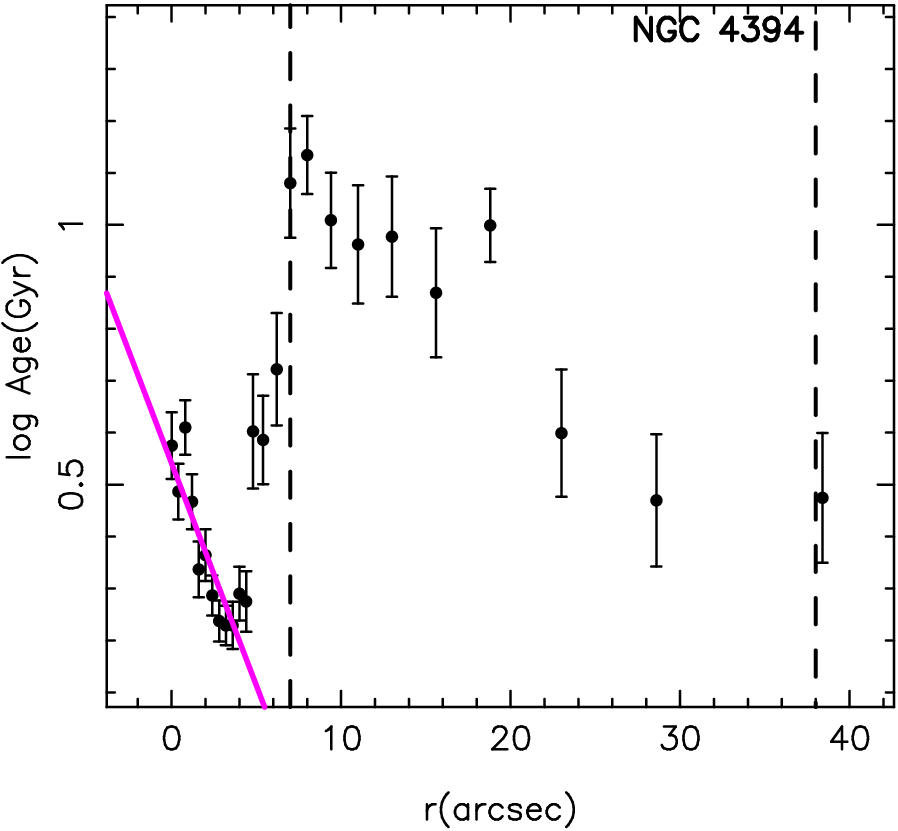}}\hspace{0.6cm}
\resizebox{0.3\textwidth}{!}{\includegraphics[angle=-90]{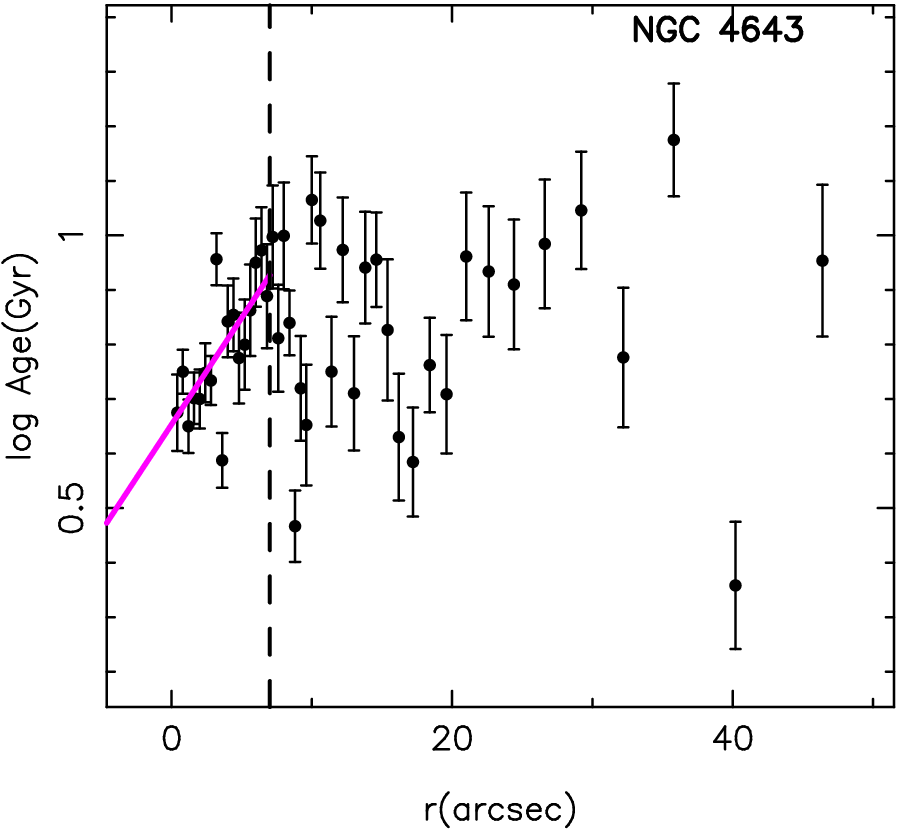}}\hspace{0.6cm}
\resizebox{0.3\textwidth}{!}{\includegraphics[angle=-90]{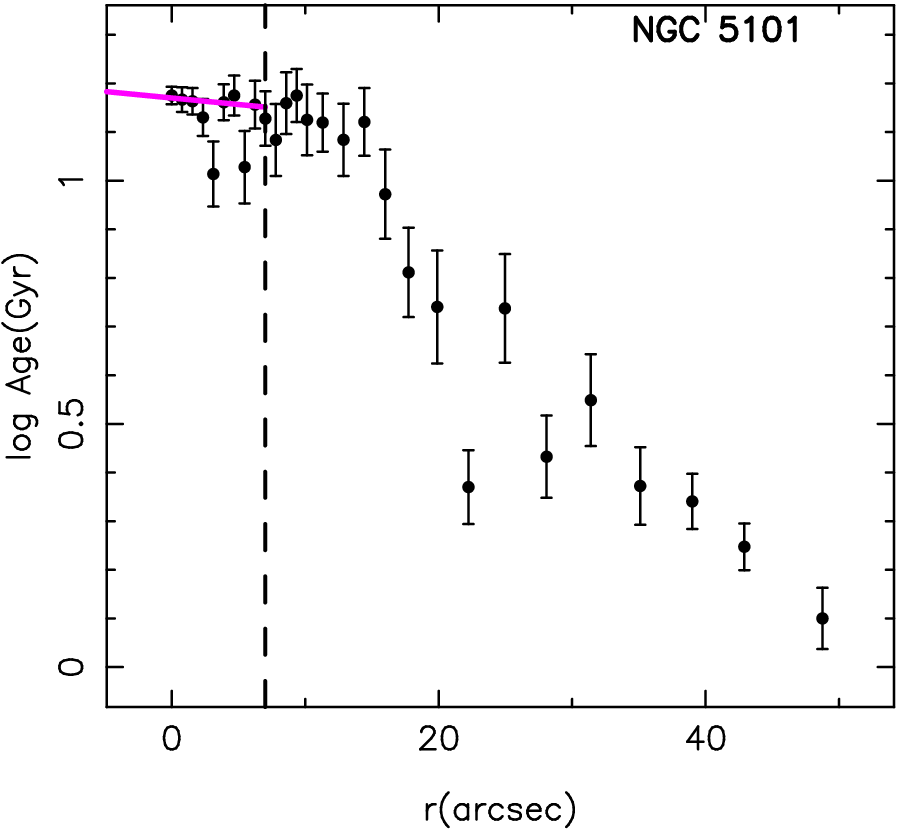}}
\caption{continued.}
 \end{figure*}
\addtocounter{figure}{-1}
\begin{figure*}
\resizebox{0.3\textwidth}{!}{\includegraphics[angle=-90]{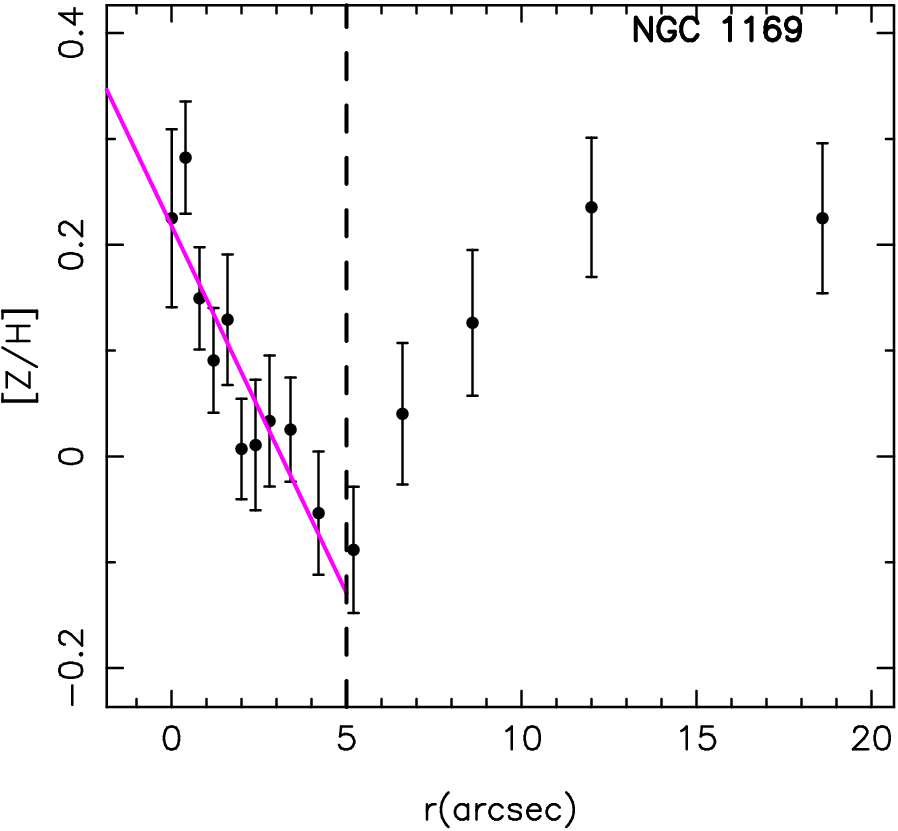}}
\resizebox{0.3\textwidth}{!}{\includegraphics[angle=-90]{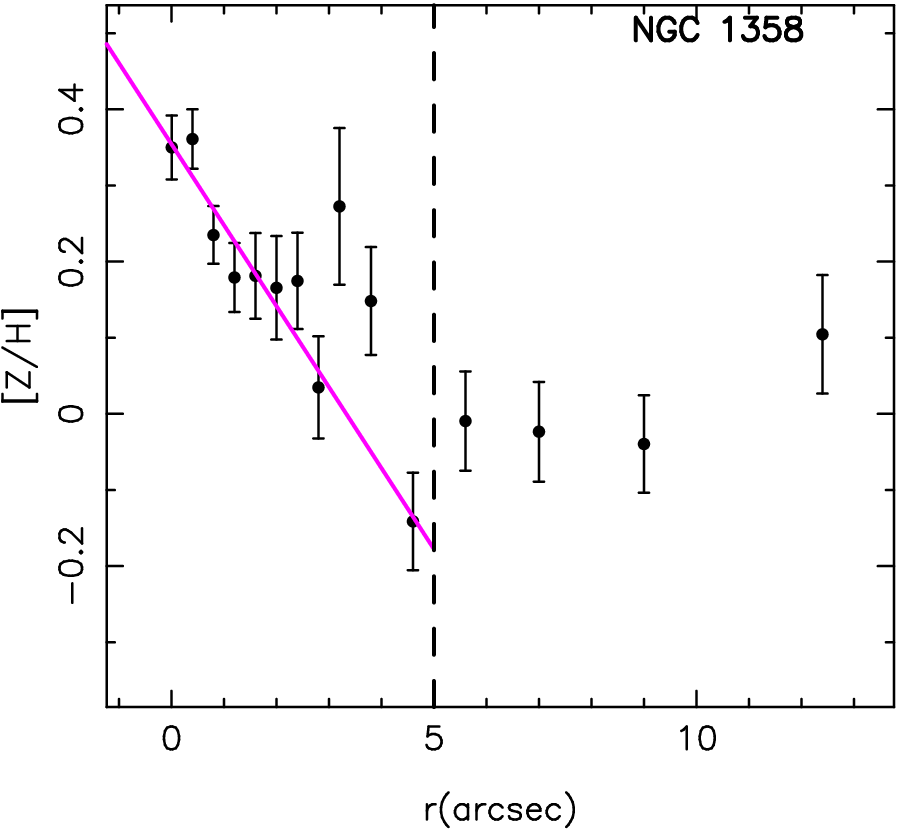}}
\resizebox{0.3\textwidth}{!}{\includegraphics[angle=-90]{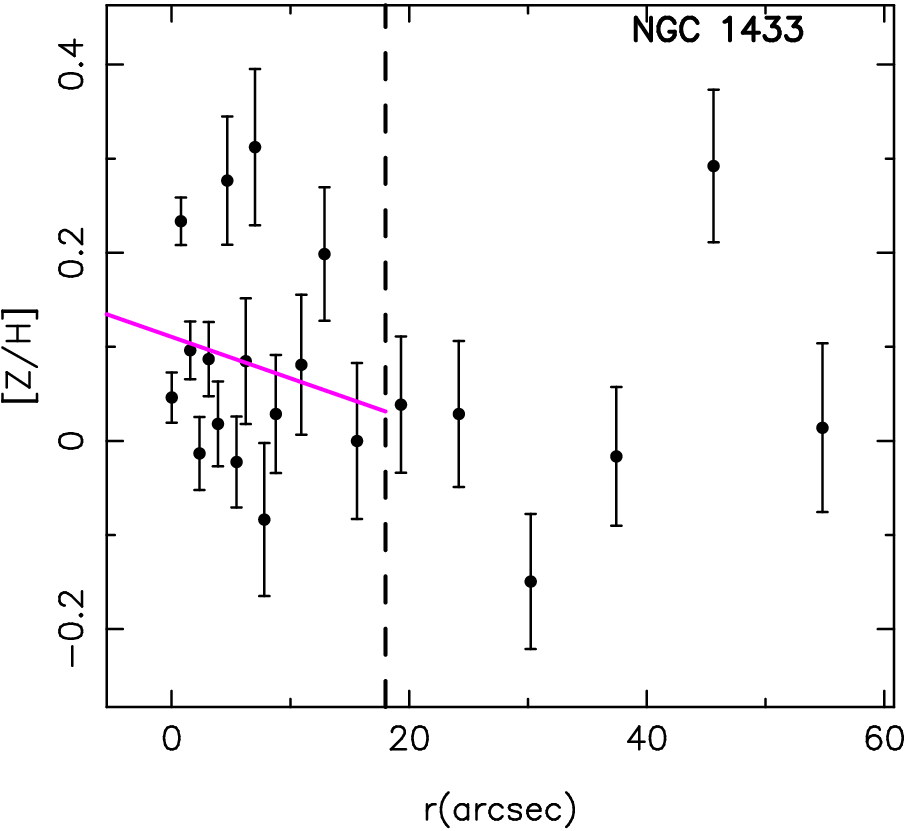}}
\resizebox{0.3\textwidth}{!}{\includegraphics[angle=-90]{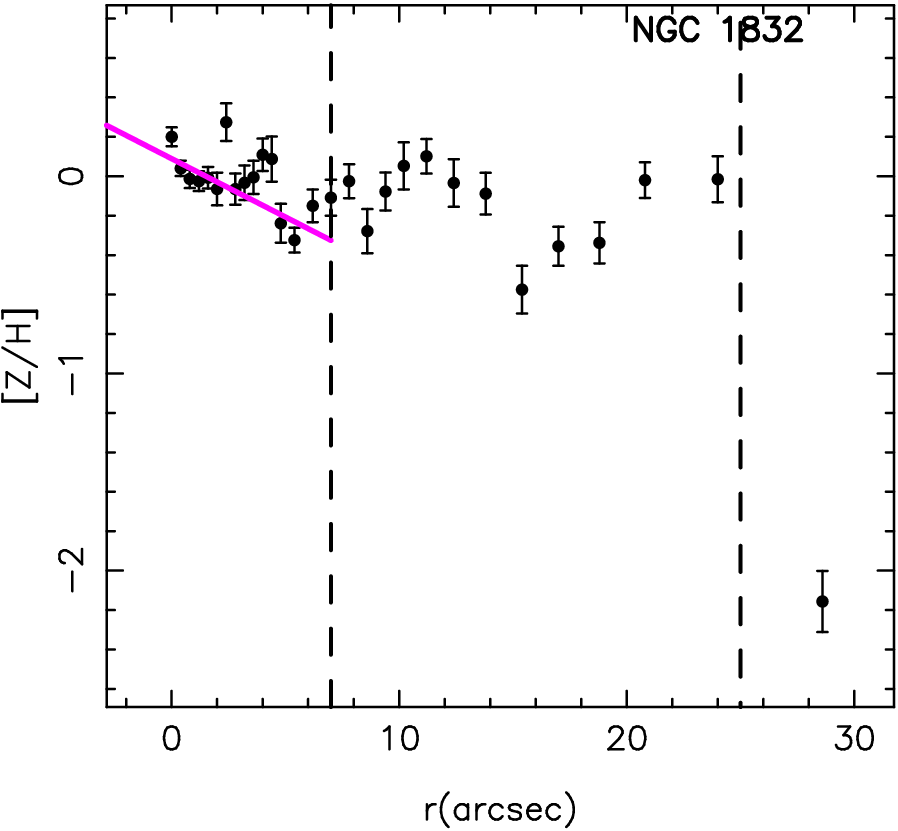}}
\resizebox{0.3\textwidth}{!}{\includegraphics[angle=-90]{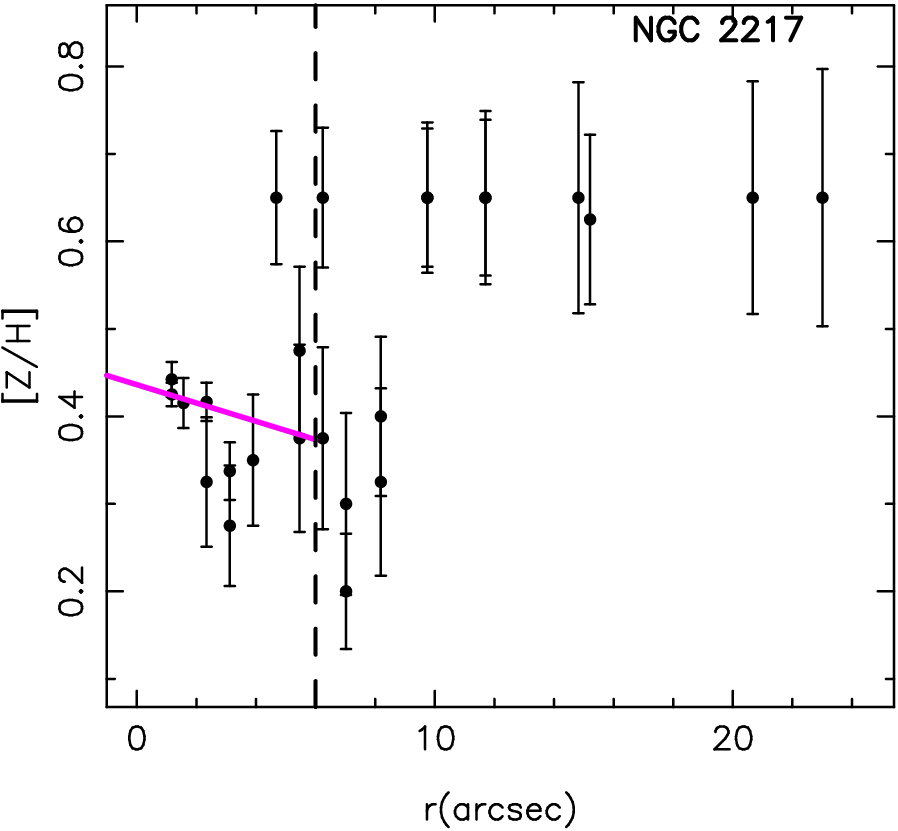}}
\resizebox{0.3\textwidth}{!}{\includegraphics[angle=-90]{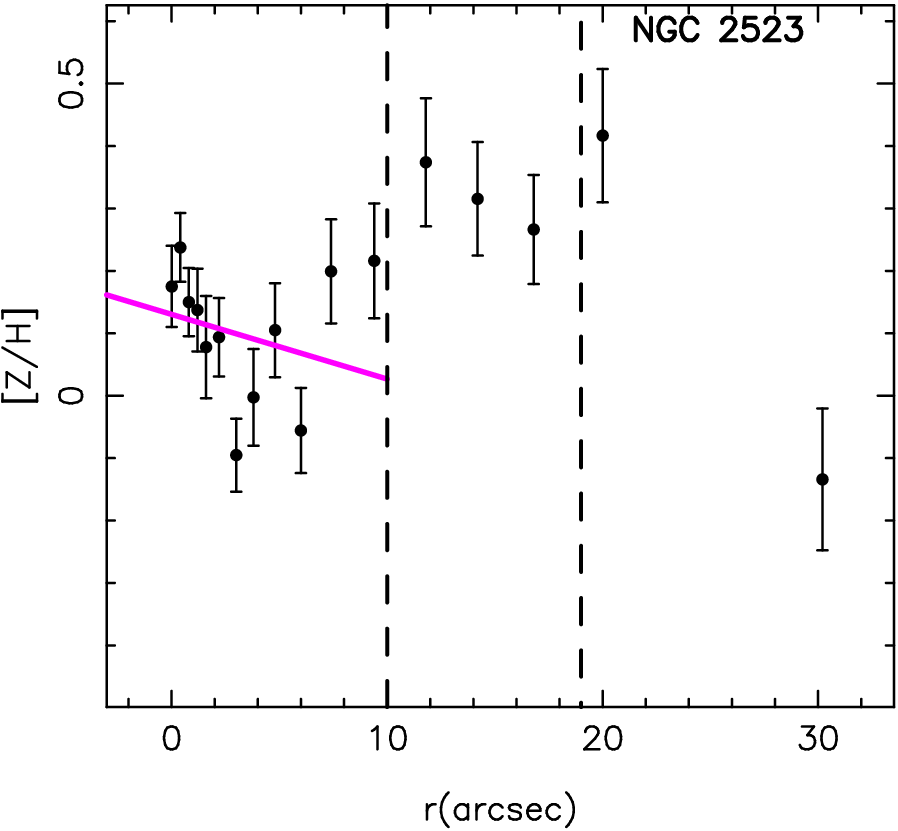}}
\resizebox{0.3\textwidth}{!}{\includegraphics[angle=-90]{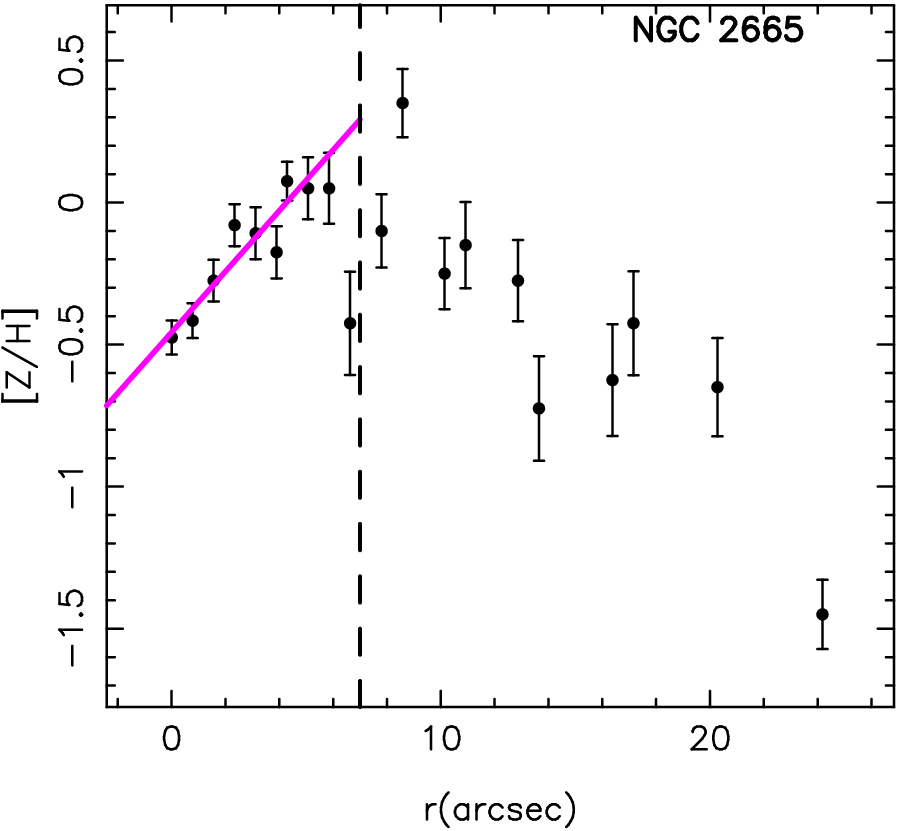}}\hspace{0.6cm}
\resizebox{0.3\textwidth}{!}{\includegraphics[angle=-90]{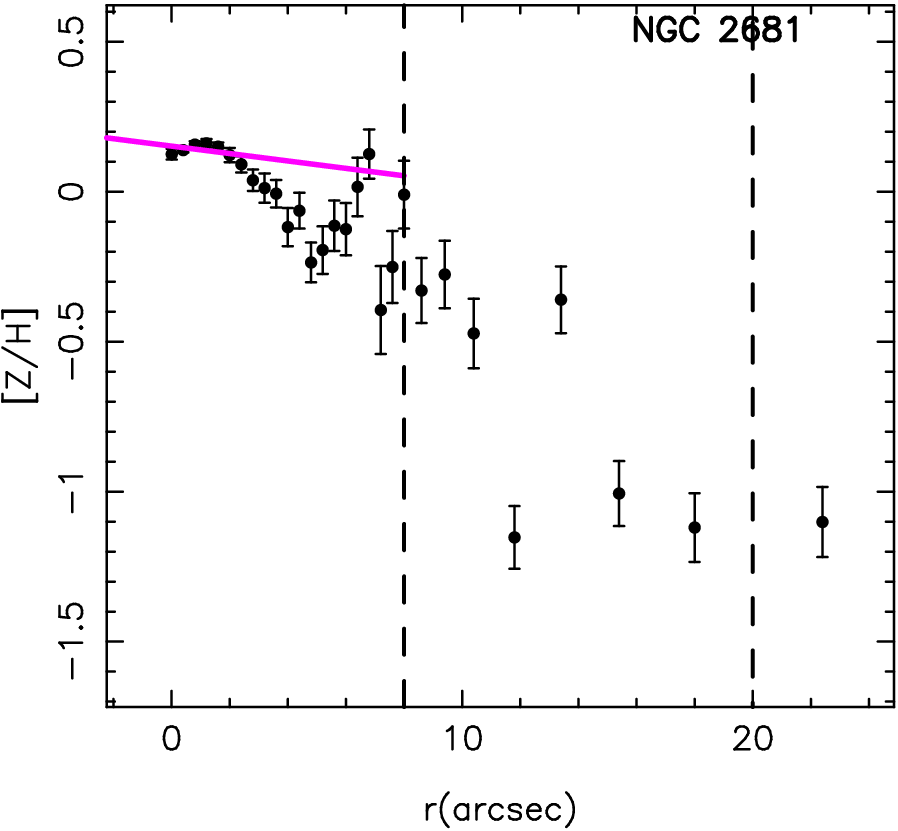}}\hspace{0.6cm}
\resizebox{0.3\textwidth}{!}{\includegraphics[angle=-90]{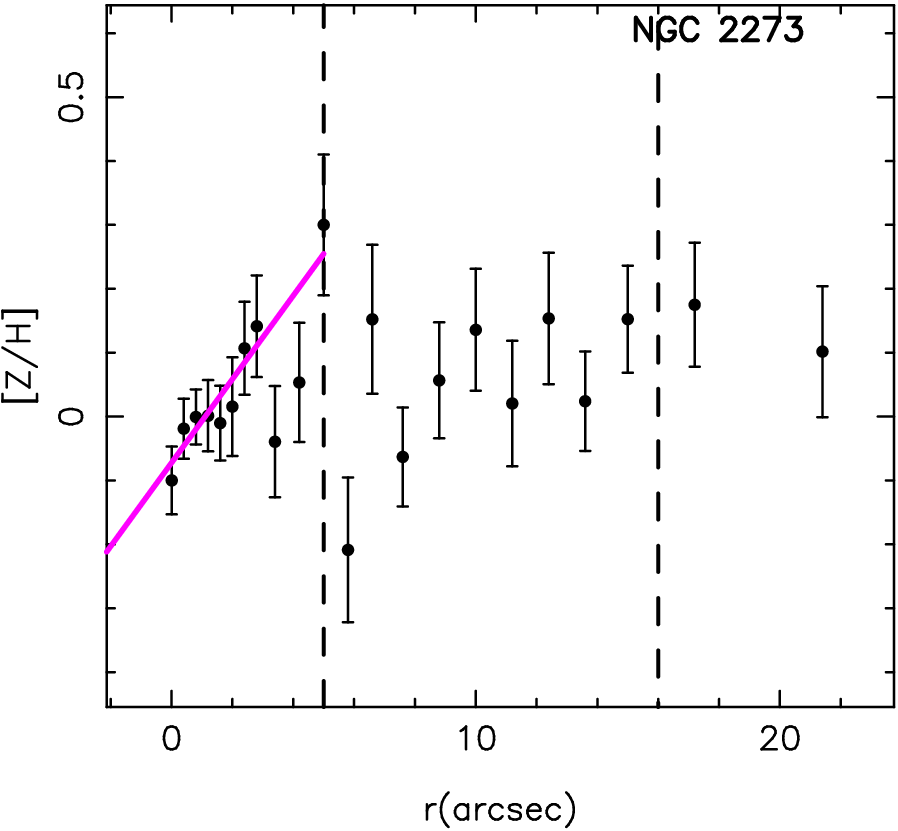}}
\caption{continued.}
\end{figure*}
\addtocounter{figure}{-1}
\begin{figure*}
\resizebox{0.3\textwidth}{!}{\includegraphics[angle=-90]{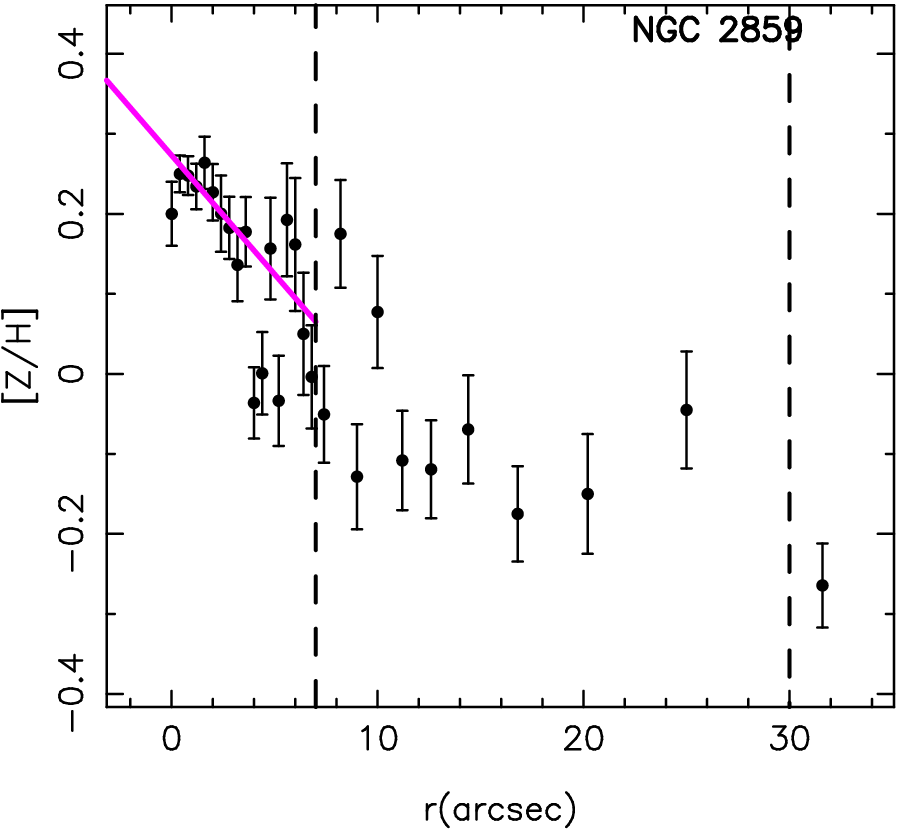}}
\resizebox{0.3\textwidth}{!}{\includegraphics[angle=-90]{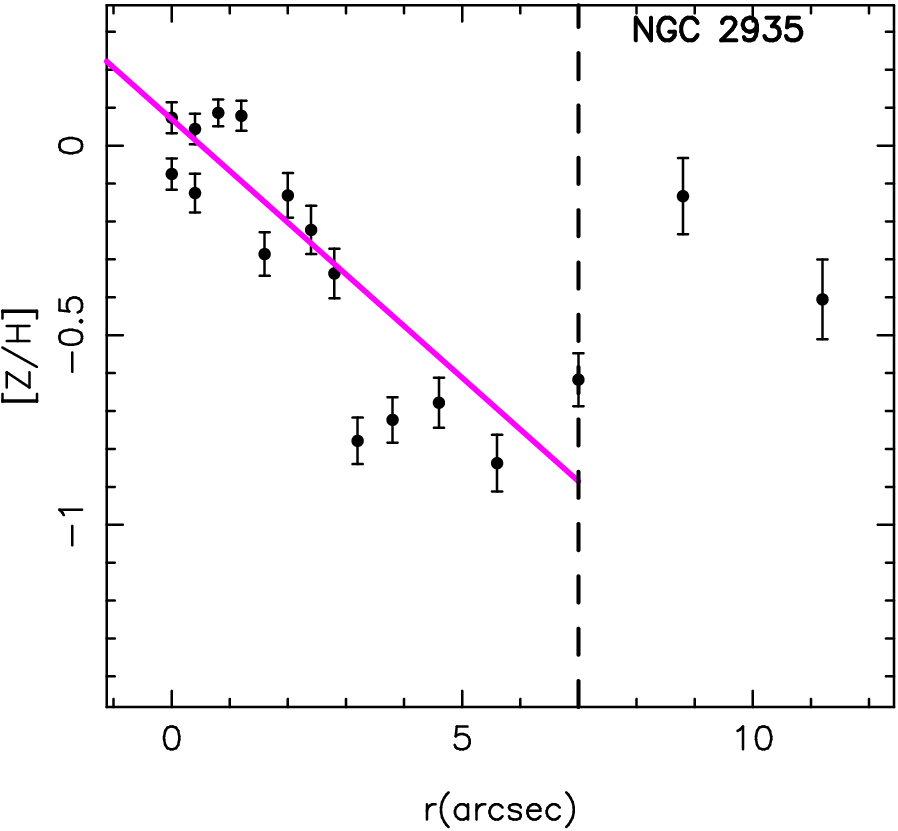}}
\resizebox{0.3\textwidth}{!}{\includegraphics[angle=-90]{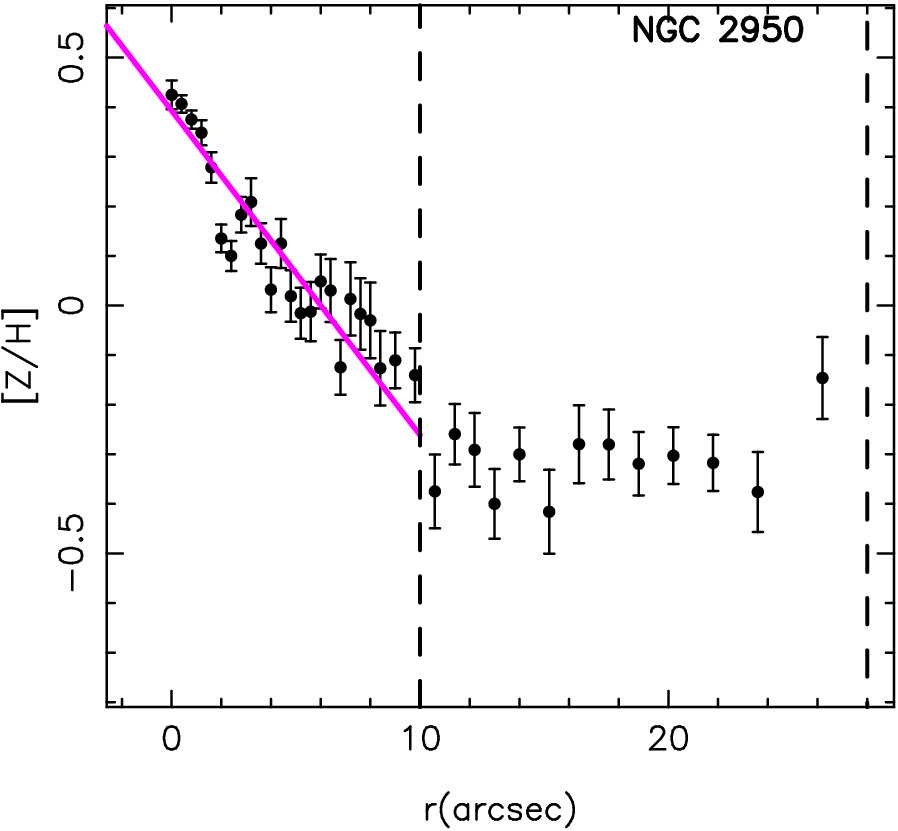}}
\resizebox{0.3\textwidth}{!}{\includegraphics[angle=-90]{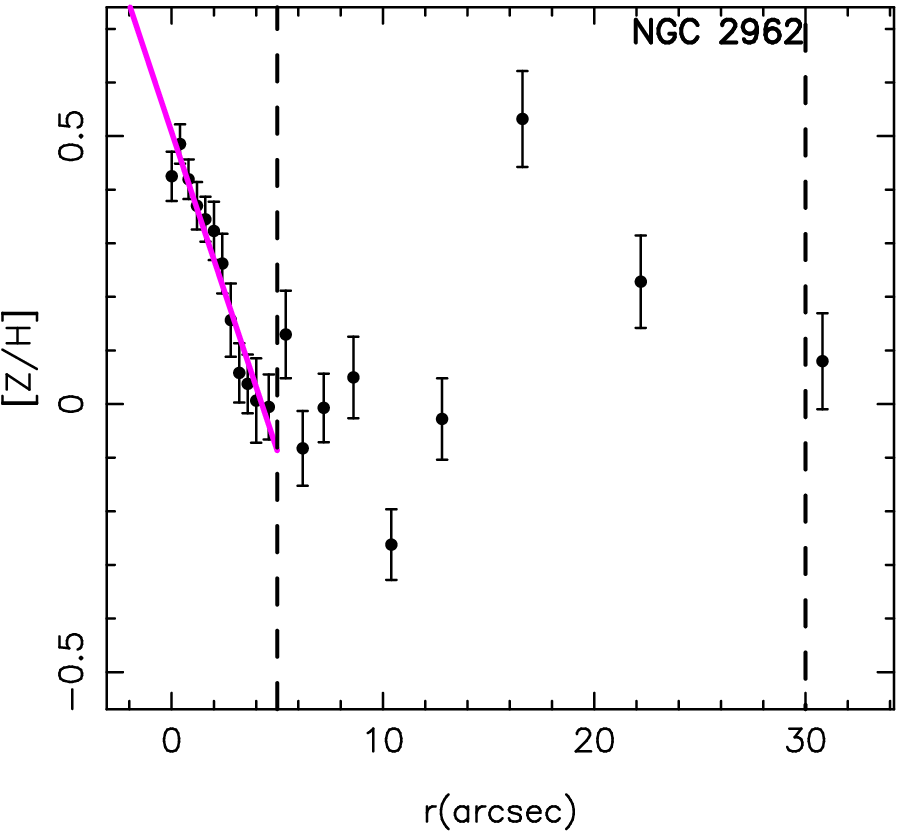}}
\resizebox{0.3\textwidth}{!}{\includegraphics[angle=-90]{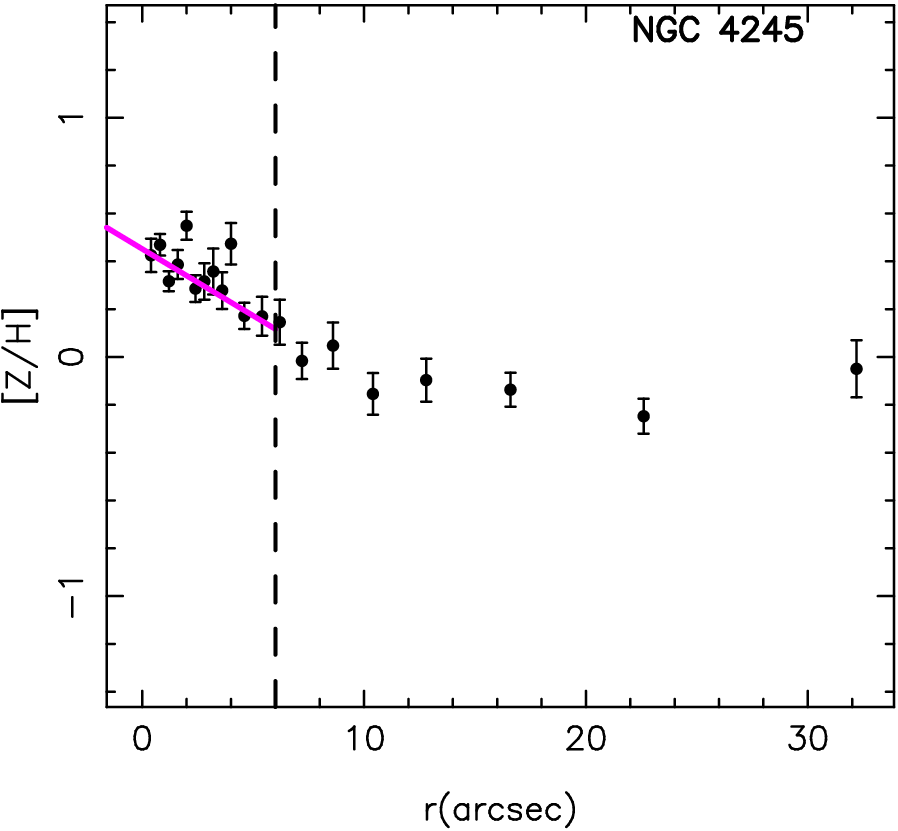}}
\resizebox{0.3\textwidth}{!}{\includegraphics[angle=-90]{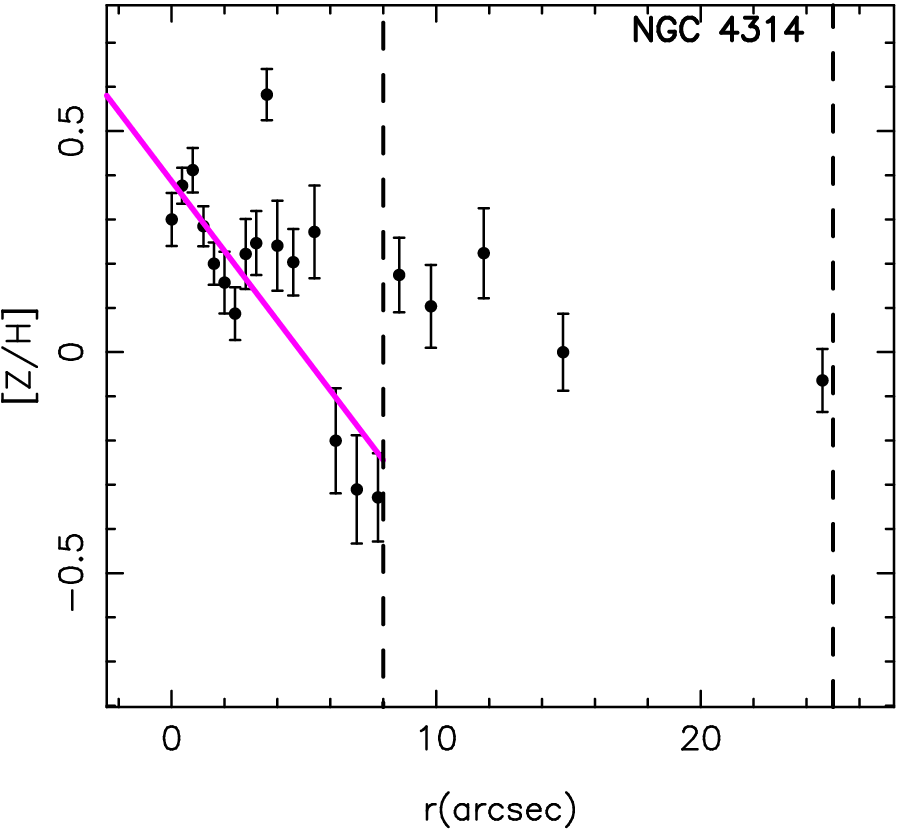}}
\resizebox{0.3\textwidth}{!}{\includegraphics[angle=-90]{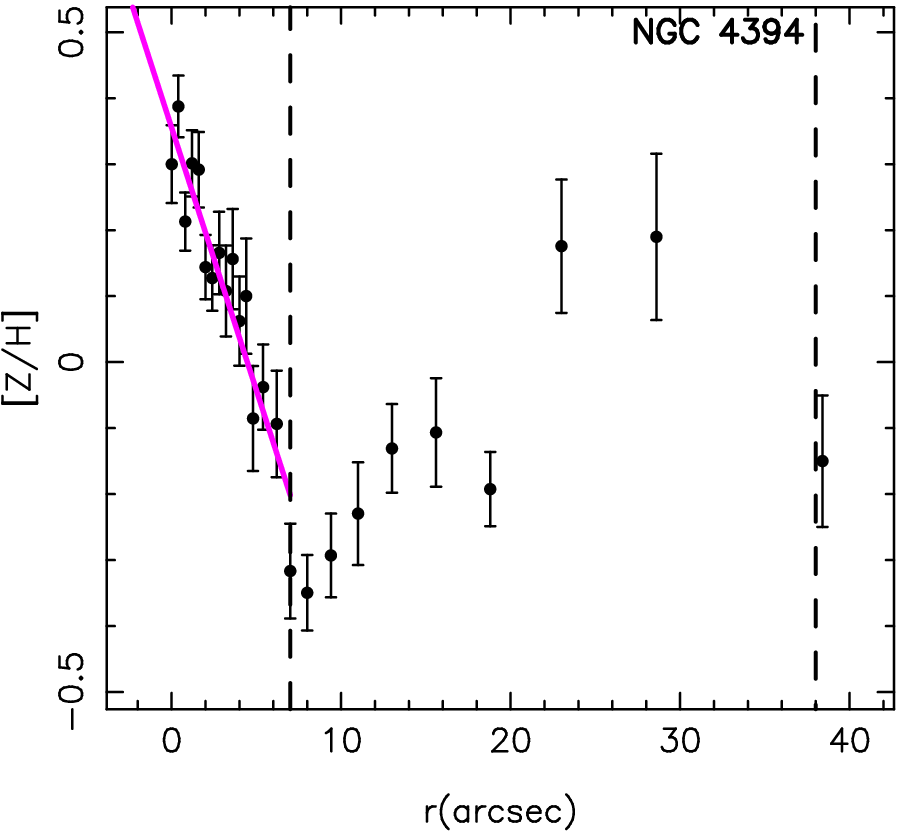}}\hspace{0.6cm}
\resizebox{0.3\textwidth}{!}{\includegraphics[angle=-90]{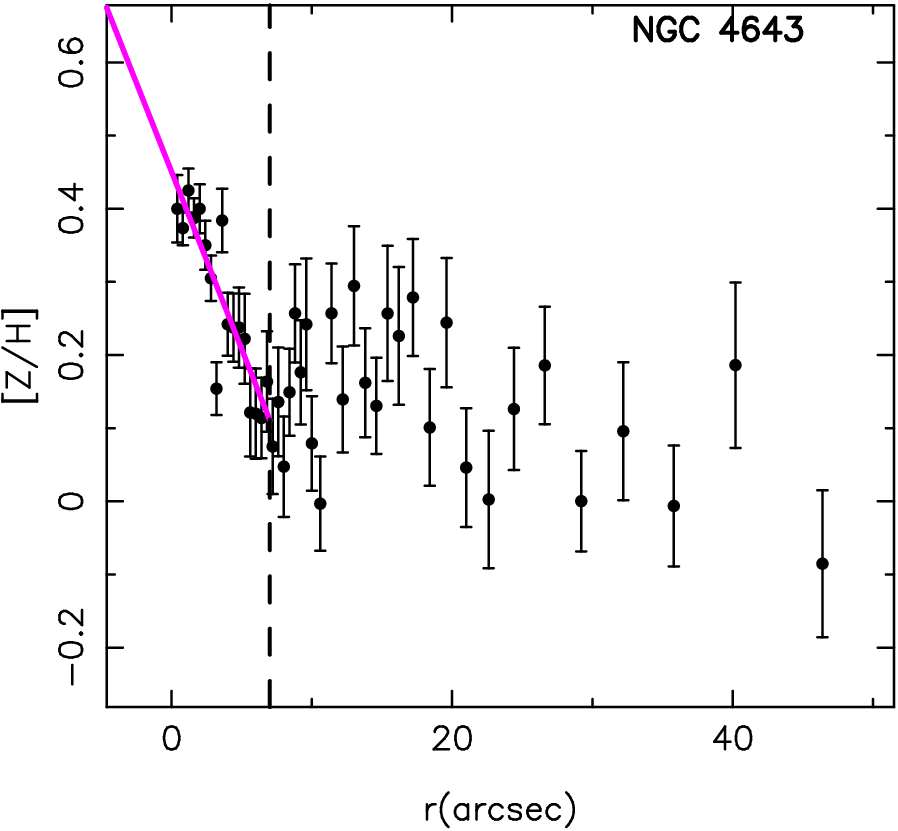}}\hspace{0.6cm}
\resizebox{0.3\textwidth}{!}{\includegraphics[angle=-90]{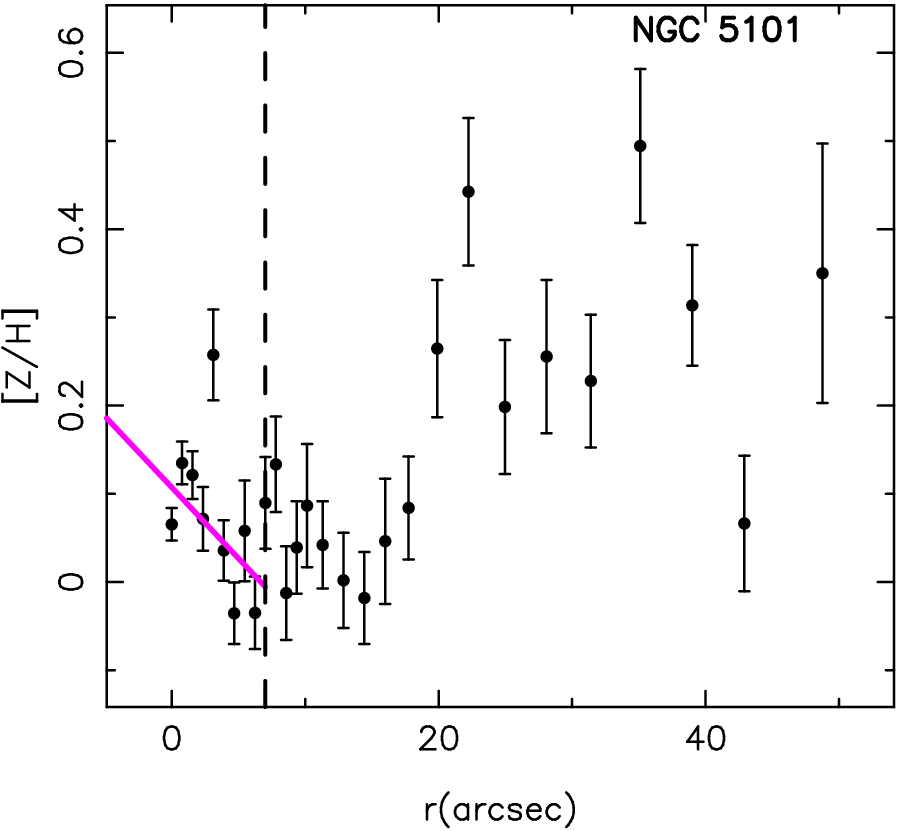}}
\caption{continued. \label{dist-age-met}}
 \end{figure*}

\begin{figure*}
 \begin{center}
\hspace{-1cm}\includegraphics[scale=0.7]{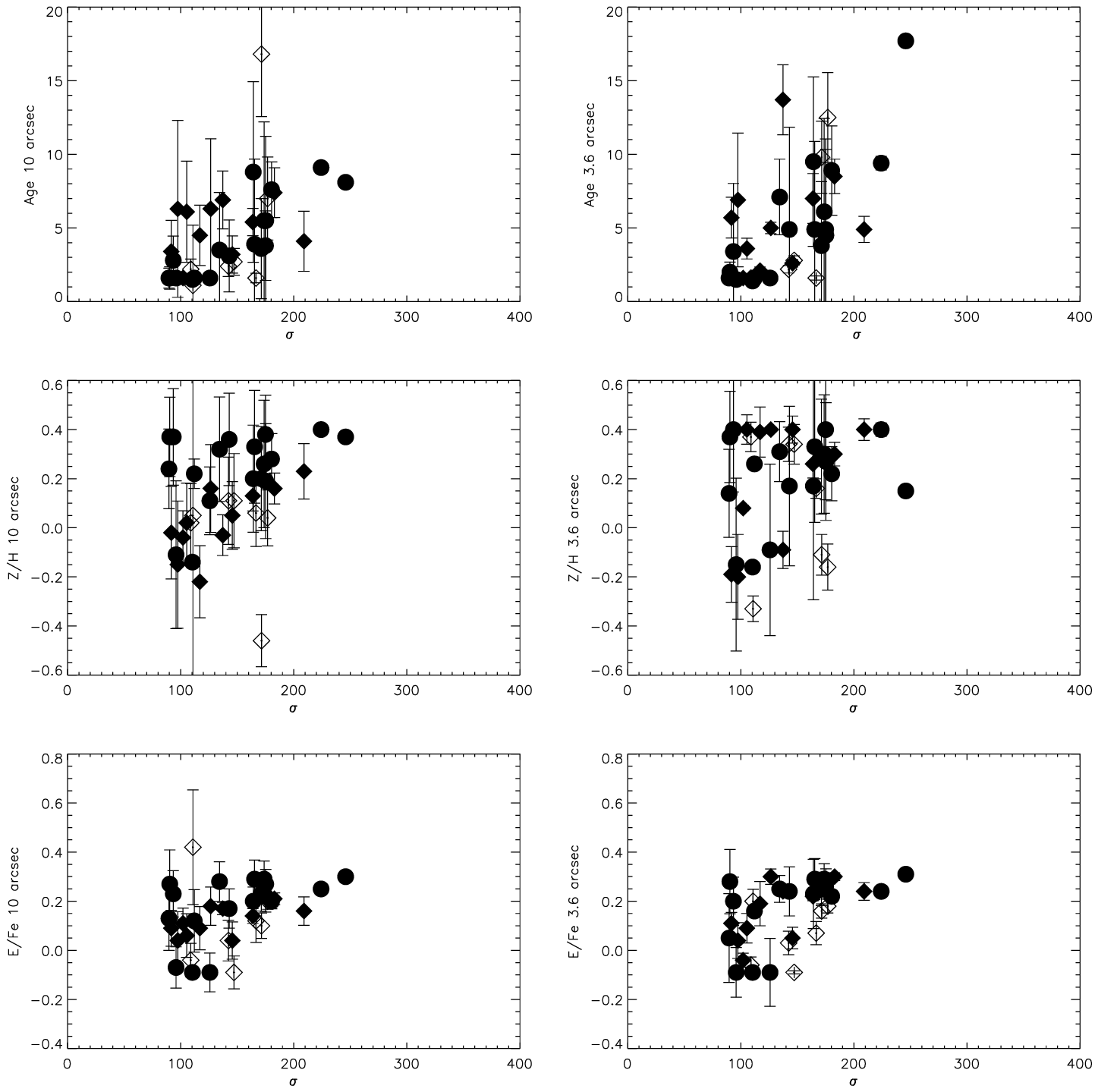}
       \caption{\label{agemet} Age, [Z/H] and [E/Fe] for the average central 10 and 3.6 arcsec apertures. Symbols as in Fig.~\ref{sigma-1.2}  }
      \end{center}
   \end{figure*}

The SSP-equivalent age and metallicity distributions along the radius are shown in Fig.~\ref{dist-age-met}. As well as for our data, we have also derived the ages and metallicities in the way explained in Paper I for the data published by Moorthy et al. (2006). From Fig.~\ref{agemet} it can be seen that the metallicity and ages in both barred and unbarred galaxies seem to cover the same range of values, but barred galaxies tend to have higher metallicities than unbarred galaxies at a given $\sigma$. No clear difference is found for the central ages. The [E/Fe] values for unbarred galaxies tend to lie below the values of barred galaxies at a given $\sigma$. When we open the aperture, to include the whole bulge, the trends become clearer, with the metallicities of the unbarred galaxies lying below those of the barred galaxies at a given $\sigma$.  Also the [E/Fe] is still remains lower for the unbarred galaxies when the aperture is as large as the whole bulge region. Similar results are found  when comparing with elliptical galaxies (see Fig.~\ref{sigma-mg} and Sect.~\ref{central.line.strengths}) although in that case, we have just compared the  Mgb~/~Fe5015 ratio.  However, these trends are week, and a significance test applied on the stellar parameters does not reveal any statistically significant difference. Although the low number of unbarred galaxies might be driving the significance test.  At this point, due to the large dispersion of the values, and the lack of statistical significance, this apparent trend on the [E/Fe]  values should be confirmed in the future with a larger set of data. The results on the bulge metallicity  confirm  the bulge spectral indices trends (see Sect.~\ref{central.line.strengths}) where we found that the metallicity sensitive indices of barred galaxies are larger than those of unbarred galaxies.  

\subsection{Age and metallicity gradients \label{age.and.metallicity.gradients}}

\begin{figure*}
 \begin{center}
\resizebox{0.45\textwidth}{!}{\includegraphics[angle=-0]{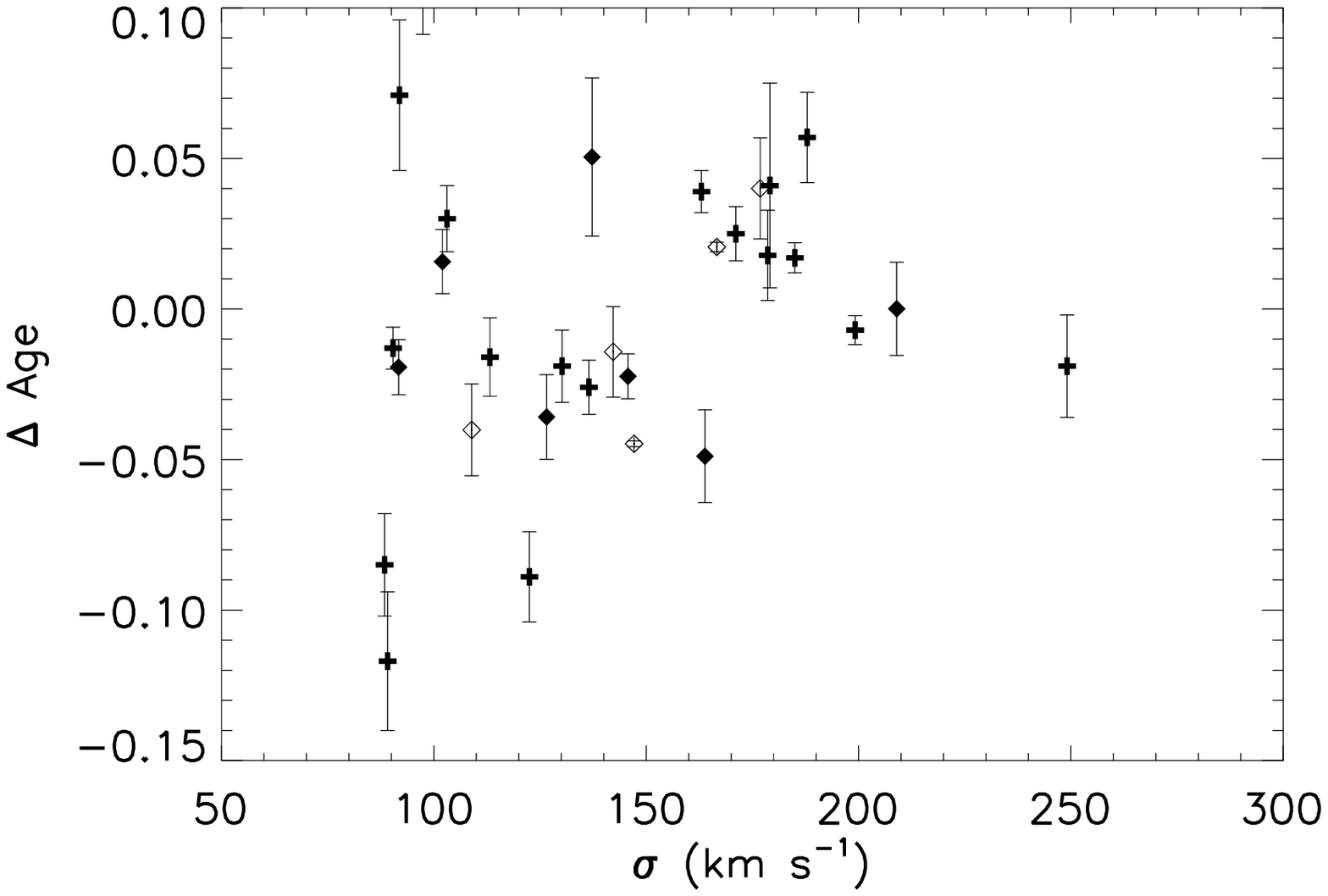}}
\resizebox{0.45\textwidth}{!}{\includegraphics[angle=-0]{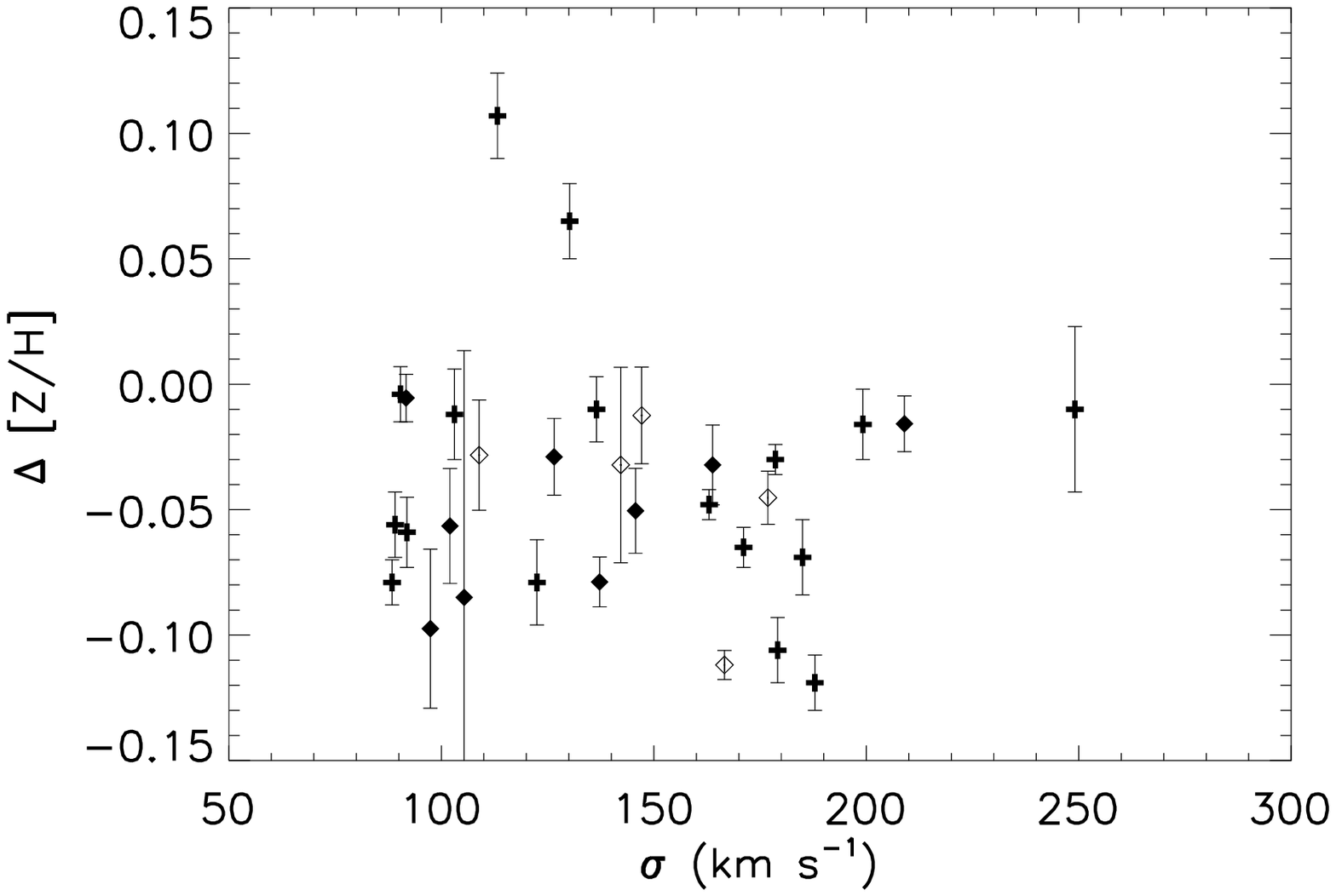}}
\resizebox{0.45\textwidth}{!}{\includegraphics[angle=-0]{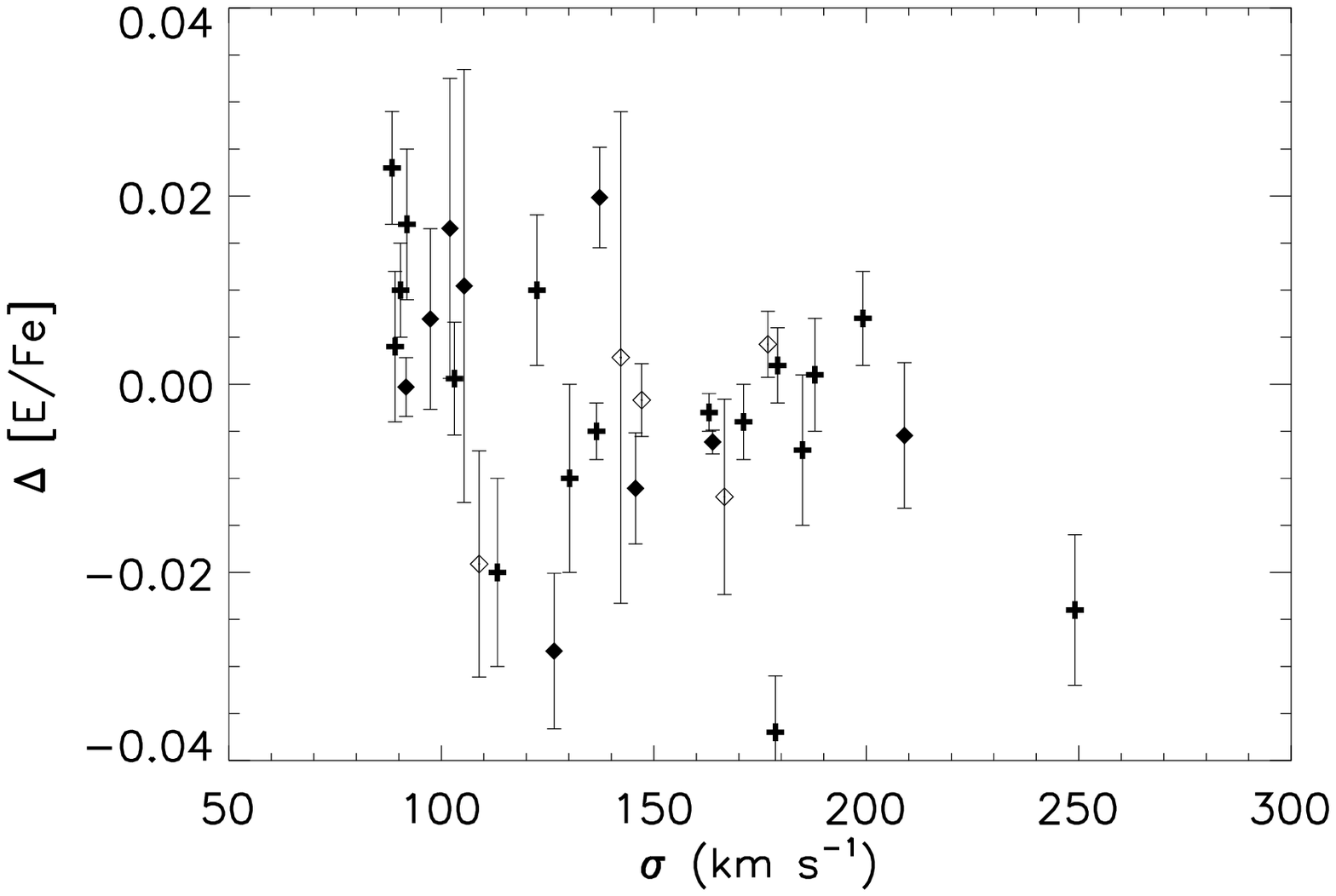}}
       \caption{\label{gradage} Age, [Z/H]  and  [E/Fe] gradient in the bulge region vs. maximum central dispersion. Symbols as previous figures, i.e empty symbols representing unbarred galaxies, while filled symbols represent barred galaxies with the crosses being our galaxies and the diamonds represent Moorthy's galaxies for which a reliable gradient could be obtained.  }
      \end{center}
   \end{figure*}

Linear fits have been performed on the derived ages, metallicities and [E/Fe] values vs. radius (see Table~\ref{grads}). Figure~\ref{dist-age-met} shows the radial distribution of ages and metallicities for all the galaxies. Fig.~\ref{gradage} shows that, on average, galaxies with lower $\sigma$ tend to have negative age gradients while larger central velocity dispersion galaxies show positive age gradients. The relation between [Z/H] gradient and the central velocity dispersion is shown in Fig.~\ref{gradage}. Most galaxies show a negative [Z/H] gradient. The mean metallicity gradient is -0.19 $\pm$ 0.07 dex for the barred galaxies and -0.3$\pm$  0.1 for the non-barred ones (notice that these gradients are calculated using log scale for the radius to be able to compare with the literature, while the gradients given in Table~\ref{grads} and Fig.~\ref{gradage} have been derived using a linear scale). Both values are within the values obtained by other authors for the bulge of galaxies. For example,
Morelli et al.  found a mean value of (-0.15 dex) and Jablonka et al. (2007), d[Z/H]/dr=-0.2 dex for a sample containing both, barred and non-barred galaxies.

The differences between the metallicity gradient of barred and non-barred galaxies are not statistically significant. We also do not 
find a correlation between the metallicity gradient of barred galaxies and the central velocity dispersion.  In early-type galaxies, a possible correlation between the metallicity  gradient and the mass for elliptical galaxies with central sigma $\leq$150-175 km/s has been claimed (S\'anchez-Bl\'azquez et al. 2007; Spolaor et al. 2009; Kuntschner et al. 2010)\nocite{patri2007,spolaor}. Other studies, for both, early-type galaxies (Ogando et al. 2005 ) and bulges (Jablonka et al. 2007)\nocite{ogando2005} found a lower boundary instead of a correlation in the sense that massive galaxies do not  show strong gradients but less massive galaxies show a larger scatter. In our sample we do not observe neither a correlation or a lower boundary and, after comparison with the data from Jablonka et al. 2007, our gradients are distributed similarly to their gradients and therefore, there is no systematic effect affecting our results. Very high S/N data might help to answer whether there is a real correlation between metallicity gradients and velocity dispersion.
[E/Fe] gradients are shown in Fig~\ref{gradage}. The [E/Fe] gradients are very small, almost compatible with being null, although they take both positive and negative values. Bulges with low velocity dispersion ($\sigma$ $<$ 100km\,s$^{-1}$) tend to have systematically positive values, while the same is not true for more massive bulges. In any case, the values of the gradients are, as said above, very small. This is compatible with what is found in elliptical galaxies, where also the [E/Fe] is found to be compatible with zero in most cases (e.g., Mehlert et al. 2003; Sanchez-Blazquez et al. 2006, 2007)\nocite{mehlert} and in other studies of bulges (e.g., Jablonka et al. 2007; MacArthur et al. 2009).  For comparison, we have also derived, in a similar way, age, metallicity and [E/Fe] gradients for Moorthy's sample. The values of the barred galaxies in this sample coincide with those of our galaxies and we do not find difference between the values of barred and unbarred galaxies. The interpretation of the gradients is complicated as their value may change due to many different physical processes and also because substructures inside the galaxy, as rings, disks, etc, which are common in our sample, change the value of the slope which is intrinsic to the spheroid. 
      
\begin{table*}
\caption{ Linear fit age and metallicity gradients to the bulge region}
\label{grads}
\begin{tabular}{l r r r r r r r r}     
\hline \hline
Name & Age grad&  err & [Fe/H] grad & err & [E/Fe] & err & [Z/H] & err \\
\hline
N1169&$   0.017$&0.005&$-0.064 $&  0.013 &$-0.007$ &  0.008 &$ -0.069$& 0.015 \\
N1358&$   0.041$&0.034&$-0.105 $&  0.013 &$  0.002$ &  0.004 & $-0.106$&0.013   \\
N1433&$ -0.013$&0.007&$-0.001 $&  0.009 &$  0.010$ &  0.005 &$ -0.004$&  0.011 \\
N1832&$   0.071$&0.025&$-0.061 $&  0.020 &$  0.017$ &  0.008 & $-0.059$ &  0.014 \\
N2217&$ -0.019$&0.017&$-0.030 $&  0.019 &$-0.024$ &  0.008 &$-0.010$  & 0.033 \\
N2523&$ -0.026$&0.009&$  0.010 $&  0.006 &$-0.005$ &  0.003 & $-0.010$ &  0.013 \\
N2665&$ -0.016$&0.013&$  0.121 $&  0.028 &$-0.020$ &  0.010 &$ 0.107 $ & 0.017 \\
N2681&$   0.030$&0.011&$-0.021 $&  0.014 &$  0.001$ &  0.006 &$-0.012$   &0.018 \\
N2273&$ -0.019$&0.012&$  0.070 $&  0.018 &$-0.010$ &  0.010 &$ 0.065$   &0.015 \\
N2859&$   0.018$&0.015&$-0.037 $&  0.006 &$-0.037$ &  0.006 &$-0.030$  & 0.006 \\
N2935&$   0.098$&0.022&$-0.121 $&  0.021 &$-0.022$ &  0.010 &$ -0.136$  & 0.025 \\
N2950&$   0.025$&0.009&$-0.069 $&  0.004 &$-0.004$ &  0.004 &$-0.065$  & 0.008 \\
N2962&$   0.057$&0.015&$-0.105 $&  0.009 &$  0.001$ &  0.006 &$ -0.119$  &0.011 \\
N4245&$ -0.117$&0.023&$ -0.025$&  0.022 &$  0.004$ &  0.008 &$-0.056$ &  0.013 \\
N4314&$ -0.089$&0.015&$ -0.057$&  0.020 &$  0.010$ &  0.008 &$-0.079$   &0.017 \\
N4394&$ -0.085$&0.017&$ -0.087$&  0.011 &$  0.023$ &  0.006 & $-0.079$   &0.009 \\
N4643&$   0.039$&0.007&$ -0.043$&  0.006 &$-0.003$ &  0.002 &$-0.048$   & 0.006 \\
N5101&$ -0.007$&0.005&$ -0.026$&  0.013 &$  0.007$ &  0.005 & $-0.016$ &  0.014 \\
\hline
  
\end{tabular}
\end{table*}
 
\subsection {Comparison between the SSP stellar parameters in the bulge and the bar region \label{comparison.bar.bulge}}
\begin{figure}
 \begin{center}
\hspace{-1cm}\includegraphics[width=8.0cm]{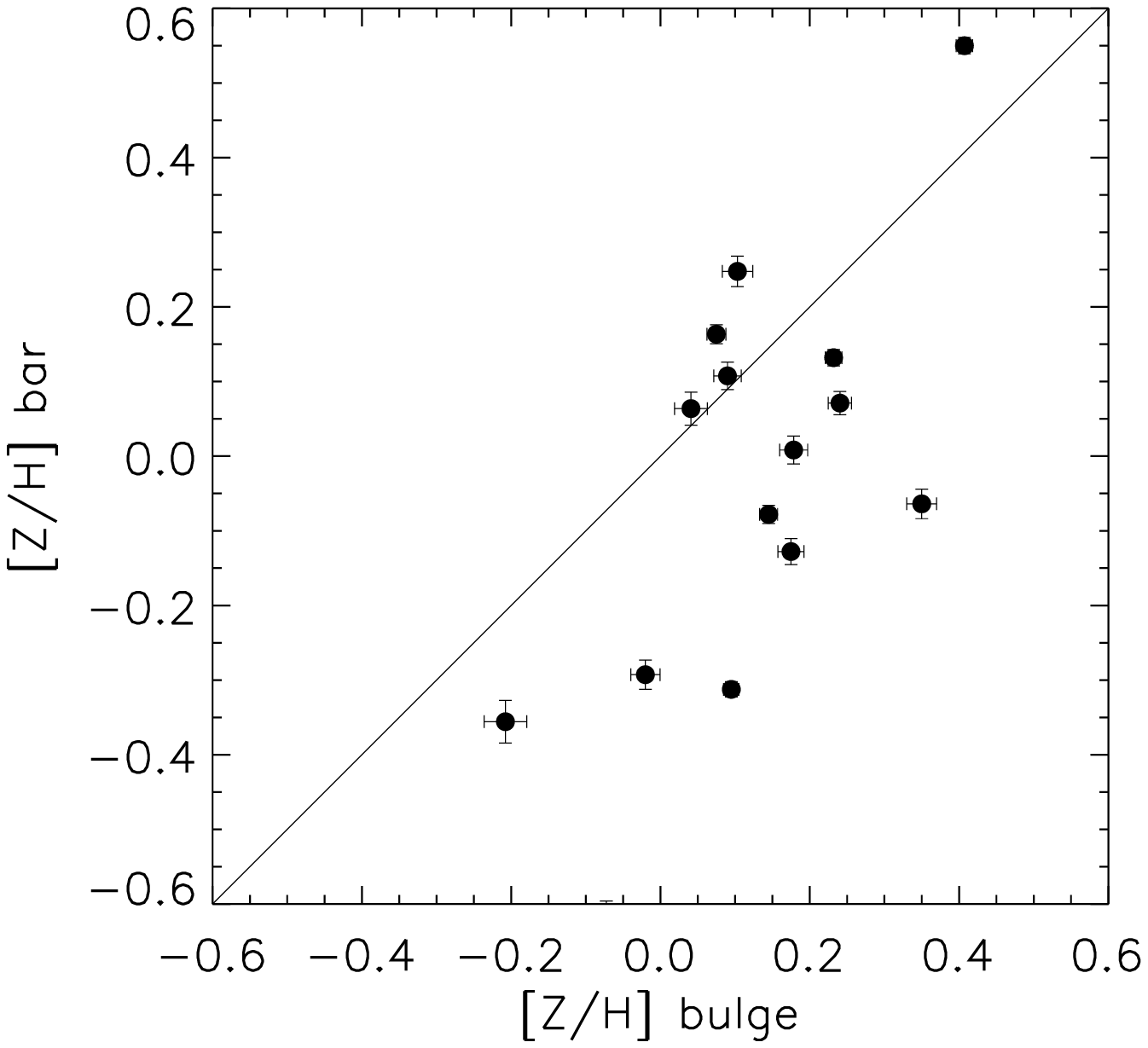}
       \caption{ \label{comparison}  Average metallicity values of the bulge and the bar region (from Paper I) for the sample galaxies.}
      \end{center}
   \end{figure}

It is interesting to relate the bulge mean age and metallicity gradients to the gradients and mean values in the bar region because in Paper I, we found seven galaxies with metallicity gradients  in the bar region significantly (more than 3-$\sigma$ significance) different from zero:
NGC~1169, NGC~2217, NGC~4394,  and NGC~5101 (positive) NGC~2665, NGC~2681, NGC~4245 (negative), we excluded from this analysis NGC~1530, NGC~3081, NGC~4314 and NGC~2935 due to the large  fitting errors in the bar region.   
In Fig.~\ref{comparison} we can see the relation between the mean metallicities in the bulge and the bar region.  It is striking the good correlation between the metallicities in both the bar and the bulge. This results points to a bulge enrichment linked to that of the bar. 

Although the mean values of both bars and bulges are correlated,  the bulge gradients do not show any correlation with the bar gradients, being the bulge gradients more indicative of the presence of morphological substructures in the bulge. 

\section{Summary and discussion\label{discussion}}

Following the results from P\'erez et al. (2009)  for the stellar line-strength indices and the ages and metallicities distribution along the bars of a sample of 20 galaxies, we have carried out a detailed study of the populations of the central regions of those galaxies. We have compared the results with the bulge properties of a similar sample of unbarred galaxies (Moorthy et al. 2006), deriving the SSP equivalent stellar parameters in the same way as for our sample. We have found differences in the bulge stellar population properties between barred and unbarred galaxies. We find that some of the metallicity sensitive indices of the bulges of barred galaxies lie above those of unbarred galaxies and the Balmer indices tend to follow the opposite trend, as was already hinted by Moorthy et al. (2006). It is interesting to note that although one could say that there is a good correlation between line-strength indices and $\sigma$,  it would be a better description of the index distribution with $\sigma$  to say that there is a almost constant value of the indices with $\sigma$ plus a tail in the distribution for the smaller $\sigma$ galaxies (below $\approx$150km~\,s$^{-1}$). Most galaxies show negative gradients for metal-sensitive indices while the opposite  is true for Balmer line indices, although NGC~1832 and NGC~2665 show clearly the opposite trend. The index distributions are closely linked to morphological substructures in the bulge region such as nuclear rings.

We have derived the ages, metallicities and [E/Fe] values using two different set of models. The central parts of barred galaxies tend to be more metal rich than the unbarred counterparts. Interestingly,  [E/Fe] seems to be enhanced for the barred galaxies over the unbarred central regions. Although, due  perhaps to the low number of unbarred galaxies in the sample, these trends are not statistically significant and would need to be followed up with a larger sample of unbarred galaxies. The age gradients follow the index trend with 
$\sigma$, most galaxies show no gradient or slightly positive gradient; however, the distribution of gradients for galaxies with $\sigma$ below $\approx$ 150km~\,s$^{-1}$\ shows a larger dispersion.  The same behaviour is found for the metallicity trends, with most values being negative, with grad[Fe/H] around -0.05, the only outliers are found in galaxies with $\sigma$ below 150km~\,s$^{-1}$\ . Galaxies with $\sigma$ above 150km\,s$^{-1}$\  show a null [E/Fe] gradient while galaxies with central velocity dispersion below this value show a positive [E/Fe] gradient. We have calculated the gradients also in Moorthy's data. There is a very good agreement in the values of both samples and we do not find any difference between barred and unbarred galaxies. The interpretation of the gradients is difficult as there are too many processes that can modify their shape. Furthermore, the presence of substructures, as disks or rings, which are very common in our sample, makes very difficult to measure the real gradient of the spheroid. 

The fact that barred galaxies  show a similar age and a  lower [E/Fe] is, in principle, a puzzling result.  In a classical scenario of secular building of the bulge, the star formation extends for much longer timescales than in a merger built scenario, where star formation happens very efficiently and in short period of times. 
On the contrary, we are finding (in agreement with Moorthy et al. 2006), completely the opposite result.  
One possible bias to this study is that our sample contains, mostly, early-type spirals.  We showed, in Paper I, that the bars in these galaxies are old and have high values of E/Fe (see also Gadotti \& de Souza 2006)\nocite{gadotti}.  The epoch of  formation of these bars is probably associated with the epoch 
of formation of the bulge, which would be compatible with the strong correlation we find in the stellar parameters of both components. 
Some proposed mechanisms to form bars are, e.g., mergers and interactions. These processes can create both, the bar and the bulge, in particular in those galaxies with early type morphologies (see, e.g., Walker et al. 1996; Berentzen et al. 2004;  Peirani et al. 2009).\nocite{walker96,peirani,berentzen04} In this scenario, it could well happen  
that this mechanism created both, the bar and the bulge,  in some galaxies,  in particular those with early-type morphologies. 
During the formation of the bulge, the star formation could have been enhanced by the presence of this bar, increasing the metallicity and the [E/Fe] values of the bulge with respect to those bulges lacking this structure (e.g. P\'erez \& Freeman 2006\nocite{perez06}). Later accretion of gas can then be funneled towards the center forming nuclear discs and  rings (Emsellem et al. 2001\nocite{emsellem2001}; Wozniak et~al.\ 2003, Wozniak \& Champavert 2006\nocite{wozniak03,wozniak06}), producing substructures with low level of star formation, i.e. slightly younger (showing a lower SSP-equivalent age in our data), but  that would not contain enough stars to produce a change in the SSP-equivalent metallicities and [E/Fe] as the metallicity and [E/Fe] values reflect more 
the values of the dominant (in mass) stellar population  (see Serra \& Trager 2007), while the SSP-equivalent age are very bias towards the youngest components.  In order to test this hypothesis, further work and analysis would be needed to derive star formation histories. If this hypothesis is confirmed, it would imply that bars are old and long lasting features and that bar formation is, at least for this early type barred galaxies, closely linked to the formation of the bulge.  Due to the large dispersion in the values for some of the trends found in this study, in order  to confirm them, we will need in a future work to extend the unbarred galaxy sample and the morphological range covered by it.

\begin{acknowledgements}

We really thank B. Moorthy for kindly providing us with the radial values of all the indices  of their sample. We want to thank the anonymous referee for the very useful comments and discussion. This research has made use of the NASA/IPAC Extragalactic Database (NED) which is operated by the Jet Propulsion Laboratory, California Institute of Technology, under contract with the National Aeronautics and Space Administration. I. P\'erez thanks the support of a postdoctoral
 fellowship from the Netherlands Organisation for Scientific Research (NWO, Veni-Grant 639.041.511) and she is currently supported by the Spanish Plan Nacional del Espacio del Ministerio de Educaci\'on y Ciencia(via grant C-CONSOLIDER AYA 2007-67625-C02-02). She also thanks  the Junta de Andaluc\'{\i}a for support through the  FQM-108 project.
PSB is supported by the Ministerio de Ciencia e Innovaci\'on (MICINN) of Spain through the  Ramon y Cajal programme. 
She also thanks the support of the  European for a ERC under their 6th program. This work has been supported by the Programa Nacional de Astronom\'{\i}a y Astrof\'{\i}sica of th
e Spanish Ministry of Science and Innovation under the grant AYA2007-67752-C03-01.

\end{acknowledgements}
 
\bibliography{abundance.bulge}
\bibliographystyle{natbib}

\appendix
\section{Linear fits to the line--strength distribution in the bulge region} 
\begin{table}
\caption{Linear fit parameters (Y=a+bx) in the bulge region of NGC~1169}
\label{line-strength}
\begin{tabular}{l r r r r}     
\hline \hline
Index                   &   \multicolumn{1}{c}{a}  & \multicolumn{1}{c}{err a} & 
\multicolumn{1}{c}{b}  & \multicolumn{1}{c}{err b} \\ 
H$\delta_{\rm A}$  &$-3.005$ & 0.180    & $0.028$ & 0.090\\
H$\delta_{\rm F}$   &$-0.088$& 0.124  &$ 0.0140$& 0.063\\
CN$_1$                &$   0.088$&0.008     &$ -0.004$& 0.004\\
CN$_2$                &$   0.120$& 0.009    &$ -0.002$&  0.004\\
Ca4227                &$   1.271$&  0.069   &$ 0.017$&  0.034\\
G4300                  &$   5.179$& 0.163    &$ 0.197$&0.080\\
H$\gamma_A$     &$ -6.385$ & 0.210   &$-0.186$&0.103\\
H$\gamma_F$      &$ -2.036$& 0.099    &$ 0.087$&  0.048\\
Fe4383                 &$   5.480$  & 0.163 &$ 0.059$&  0.078\\
Ca4455                & $   1.308$ & 0.083  &$-0.078$  &0.040\\
Fe4531                 &$    3.507$& 0.100  &$-0.171 $&0.051\\
C4668                  &$    7.685$& 0.302   &$ -0.508$&0.146\\
H$\beta$             &$    1.621$& 0.112   &$ 0.065$ &0.054\\
Fe5015                 &$    5.671$& 0.151   &$ -0.247$&  0.073\\
Mg$_1$                 &$    0.128$& 0.003   &$ -0.004$& 0.001\\
Mg$_2$                 &$    0.287$& 0.003   &$-0.009$&  0.001\\
Mgb                      &$    4.484$& 0.073   &$-0.133$&  0.036\\
Fe5270                 &$    3.221$& 0.108    &$-0.161$ & 0.053\\
Fe5335                 &$    3.065$& 0.083    &$-0.217$ & 0.040\\
\hline
\end{tabular}
\end{table}

\begin{table}
\caption{Linear fit parameters (Y=a+bx) in the bulge region of NGC~1358}
\label{ajuste2}
\begin{tabular}{l l l l l}     
\hline \hline
Index & a  & err a & b  & err b \\ 
HdA& -1.890&0.188& 0.029&  0.093\\
HdF&  0.246&0.100& 0.142& 0.049\\
CN1&  0.082&0.002&-0.012& 0.001\\
CN2&   0.118  &0.004& -0.014& 0.002\\
Ca4227& 1.101& 0.065&  0.009& 0.033\\
G4300&   5.174& 0.115& -0.132&  0.061\\
HgA&      -6.013& 0.161& 0.239& 0.079\\
HgF&      -1.370&0.067& 0.041&  0.033\\
Fe4383&   5.563& 0.184&-0.213&  0.092\\
Ca4455&   1.247&  0.045&-0.097&  0.024\\
Fe4531&   3.500 & 0.118& -0.051&  0.058\\
C4668&     8.302& 0.229& -0.818 &0.121\\
Hbeta&     1.844 &0.064& -0.044  &0.033\\
Fe5015&    3.857&0.308 &-0.069 &0.153\\
Mg1&      0.149&  0.004& -0.006&  0.002\\
Mg2&      0.293&0.0023& -0.012&  0.001\\
Mgb&      4.561& 0.074 &-0.143  &0.038\\
Fe5270&  3.0222& 0.091&-0.107& 0.047\\
Fe5335&   2.750& 0.101 &-0.073 & 0.052\\

\hline
  
\end{tabular}
\end{table}

\begin{table}
\caption{Linear fit parameters (Y=a+bx) in the bulge region of NGC~1433}
\label{ajuste3}
\begin{tabular}{l l l l l}     
\hline \hline
Index & a  & err a & b  & err b \\ 
HdA& 0.084&0.210& 0.295 & 0.057\\
HdF& 1.188 &0.117&  0.171& 0.032\\
CN1&  0.067 &0.004& -0.008& 0.001\\
CN2& 0.093&  0.005 &-0.007&  0.001\\
Ca4227& 0.771&0.038&-0.017& 0.010\\
G4300& 4.00 &0.103& -0.099  &0.028\\
HgA& -3.378 & 0.148 &0.307& 0.045\\
HgF& -0.165& 0.066&  0.197&  0.023\\
Fe4383& 4.531&0.126& -0.149&  0.035\\
Ca4455& 1.077& 0.036&-0.020& 0.010\\
Fe4531& 3.169&0.048&-0.127&  0.016\\
Fe4668& 6.374& 0.128 &-0.161 & 0.035\\
Hbeta& 2.059&  0.075 &0.062 &0.021\\
Fe5015& 5.601 &0.136&-0.074& 0.036\\
Mg1&    0.083& 0.002& 0.004& 0.001\\
Mg2&    0.208& 0.002&  0.004&  0.001\\
Mgb&     2.921& 0.071&-0.029& 0.018\\
Fe5270&  2.741& 0.064 &-0.008& 0.017\\
Fe5335&  2.613&0.072&-0.088 &0.019\\

\hline
  
\end{tabular}
\end{table}

\begin{table}
\caption{ Linear fit parameters (Y=a+bx) in the bulge region of NGC~1530}
\label{ajuste4}
\begin{tabular}{l l l l l}     
\hline \hline
Index & a  & err a & b  & err b \\ 
 HdA& 2.173&0.210&  0.229&  0.095\\
HdF&  2.251 &0.189 &0.105 &0.086\\
CN1&  -0.044&  0.006 &-0.005&  0.003\\
CN2& -0.010&0.007&-0.004& 0.003\\
Ca4227& 0.548& 0.081& -0.073&0.037\\
G4300& 2.666 &0.151&-0.290& 0.068\\
HgA&  -1.030&0.441& 0.594& 0.202\\
HgF& 1.395& 0.195 &0.223 & 0.092\\
Fe4383& 3.488&  0.301 &-0.251&  0.140\\
Ca4455& 0.892& 0.074& -0.004&  0.036\\
Fe4531& 2.049& 0.395&-0.123 &0.184\\
Fe4668& 6.299& 0.242&-0.814 &0.115\\
Hbeta& 3.297& 0.209&0.084 &0.112\\
Fe5015& 5.232& 0.255& -0.748 &0.124\\
Mg1&   0.076 & 0.002 &-0.007&  0.001\\
Mg2&   0.186 &0.006 &-0.021&  0.003\\
Mgb&   2.732& 0.166&-0.197 &0.079\\
Fe5270& 2.613&  0.091 &-0.319&  0.044\\
Fe5335& 2.501& 0.070&-0.321& 0.035\\
\hline
  
\end{tabular}
\end{table}

\begin{table}
\caption{ Linear fit parameters (Y=a+bx) in the bulge region of NGC~1832}
\label{ajuste5}
\begin{tabular}{l l l l l}     
\hline \hline
Index & a  & err a & b  & err b \\ 
 HdA& 3.966& 0.226&-1.000& 0.131\\
 HdF&  3.322 &0.168& -0.536&  0.096\\
 CN1&  -0.081&  0.004 &0.018&  0.002\\
 CN2&   -0.047 &0.005&  0.016& 0.003\\
 Ca4227& 0.610& 0.073&  0.059 &0.040\\
 G4300&  1.769&0.247& 0.642& 0.130\\
HgA&     1.700&0.436 &-1.281&0.242\\
HgF&      2.815 &0.238&-0.737&0.131\\
Fe4383&  2.490 & 0.222& 0.4223&  0.114\\
Ca4455&  0.792 & 0.064&  0.042&  0.034\\
Fe4531&  2.432& 0.168&0.133&0.086\\
C4668&   4.478&0.240&0.108&0.120\\
Hbeta&    4.025  &0.150 &-0.382& 0.074\\
Fe5015&  4.398&0.212& 0.101 &0.106\\
Mg1&      0.044& 0.002& 0.004&  0.001\\
Mg2&       0.1340 & 0.005&  0.009&0.002\\
Mgb&       2.148&0.092& 0.206 &0.048\\
Fe5270&   2.247 & 0.101 & 0.046 &0.051\\
Fe5335&   2.058& 0.090&0.056&  0.045\\
\hline
  
\end{tabular}
\end{table}

\begin{table}
\caption{ Linear fit parameters (Y=a+bx) in the bulge region of NGC~2217}
\label{ajuste6}
\begin{tabular}{l l l l l}     
\hline \hline
Index & a  & err a & b  & err b \\ 
 HdA& -3.350&  0.123 &0.100 & 0.048\\
HdF& -0.399 &0.057& 0.006& 0.023\\
CN1&  0.203 &0.004 &-0.010 & 0.002\\
CN2&  0.234&0.004&-0.010& 0.002\\
Ca4227& 1.260&0.030 & 0.015  &0.013\\
G4300&   5.542 &0.071&-0.034& 0.029\\
HgA&     -8.555  &0.123  &0.308  &0.050\\
HgF&       -2.994&  0.075& 0.159 & 0.030\\
Fe4383&   6.391  &0.161 &-0.045  &0.067\\
Ca4455&   1.282&0.060& 0.005& 0.025\\
Fe4531&    3.551 & 0.137&-0.010 &0.057\\
C4668&    9.621&0.160& -0.372&0.066\\
Hbeta&     2.093&0.079&-0.141& 0.032\\
Fe5015&   6.554&0.117&0.018  &0.048\\
Mg1&       0.165& 0.002&  0.005& 0.001\\
Mg2&       0.334&0.002& 0.004 &0.001\\
Mgb  &      5.651 &0.057&-0.194& 0.023\\
Fe5270 &   3.178 &0.086 & 0.075&  0.034\\
Fe5335&  3.197 &0.064 &-0.067  &0.026\\
\hline
  
\end{tabular}
\end{table}

\begin{table}
\caption{ Linear fit parameters (Y=a+bx) in the bulge region of NGC~2273}
\label{ajuste7}
\begin{tabular}{l l l l l}     
\hline \hline
Index & a  & err a & b  & err b \\ 
 HdA& 4.650&  0.338& -0.567&  0.215\\
HdF&  3.604  &0.197&-0.285 &0.125\\
CN1& -0.071& 0.008&  0.006 &0.004\\
CN2& -0.035&0.008& 0.005 & 0.005\\
Ca4227& 0.271& 0.047& 0.095&  0.031\\
G4300& 1.456&0.210& 0.343&  0.126\\
HgA&    1.813&0.424& -0.656&0.254\\
HgF&      2.565&0.262& -0.315 &0.155\\
Fe4383&  2.4027&0.166&  0.384&0.098\\
Ca4455&  0.405&0.063& 0.120 &0.037\\
Fe4531 &  2.292 &0.173& 0.032&0.098\\
C4668& 3.129&0.246& 0.057&0.135\\
Hbeta&    2.930  &0.191& 0.012 &0.096\\
Fe5015&  -0.448 & 0.588 & 1.139 &0.296\\
Mg1&      0.101&0.008 &-0.011 &0.003\\
Mg2 &     0.141  &0.006 & 0.006& 0.002\\
Mgb &     2.150&0.099&0.143 &0.056\\
Fe5270&  1.716  &0.089 & 0.169 & 0.048\\
Fe5335 &  1.909& 0.081 &0.060&  0.043\\
\hline
  
\end{tabular}
\end{table}

\begin{table}
\caption{ Linear fit parameters (Y=a+bx) in the bulge region of NGC~2523}
\label{ajuste8}
\begin{tabular}{l l l l l}     
\hline \hline
Index & a  & err a & b  & err b \\ 
 HdA& -1.350& 0.205&0.052&0.055\\
 HdF&  0.588 & 0.106 &0.040&0.029\\
CN1&   0.029& 0.006 &-0.003&  0.002\\
  CN2&  0.058  &0.006& -0.002&0.002\\
 Ca4227& 1.158&0.073&-0.013&0.020\\
 G4300&  5.034  &0.129 &-0.072&0.035\\
 HgA&      -4.973&0.187&0.072  &0.049\\
 HgF&    -0.870 & 0.076 & 0.032  &0.020 \\
 Fe4383&  4.684& 0.134 &0.114 &0.034\\
Ca4455&   0.914 & 0.089 &0.049&  0.022 \\
 Fe4531&   3.104 &0.133&0.023 &0.033\\
 C4668&   6.484  &0.286&-0.017  &0.070 \\
 Hbeta&    1.918&0.087 &0.007 & 0.021\\
Fe5015 &    5.203 & 0.183&-0.028  &0.044\\
 Mg1 &     0.087& 0.003 &-0.000&  0.001\\
 Mg2&       0.227  &0.003&-0.001&  0.001\\
 Mgb&      3.825& 0.085 &-0.040  &0.021\\
 Fe5270&  2.870&  0.090&-0.038&  0.022\\
 Fe5335&   2.581&0.100&-0.031  &0.024\\

 \hline
  
\end{tabular}
\end{table}

\begin{table}
\caption{ Linear fit parameters (Y=a+bx) in the bulge region of NGC~2665}
\label{ajuste9}
\begin{tabular}{l l l l l}     
\hline \hline
Index & a  & err a & b  & err b \\ 
 HdA   &         3.632  &   0.243 & -0.062& 0.095\\
HdF    &        2.590 &  0.197 &0.059& 0.077\\
CN1    &       -0.027 &  0.005& 0.002& 0.002\\
CN2    &       -0.012&  0.006& 0.006& 0.002\\
Ca4227&      0.289  & 0.044 & 0.022& 0.017\\
G4300    &     0.850& 0.140& 0.318  &0.055\\
HgA      &      2.402 & 0.283& -0.421& 0.113\\
HgF     &        2.647& 0.180 & -0.216& 0.071\\
Fe4383 &       2.010& 0.248& 0.159& 0.097\\ 
Ca4455   &     0.397 & 0.095 &  0.075&  0.037\\
Fe4531  &       1.417& 0.101 & 0.146& 0.040\\
C4668   &      2.514& 0.217& 0.259& 0.083\\
Hbeta     &      3.728& 0.153& -0.246 & 0.056\\
Fe5015   &     2.998 & 0.120& 0.115& 0.047\\
Mg1    &        0.0426& 0.006  &0.011& 0.002\\
Mg2     &       0.107& 0.006 & 0.016 & 0.002\\
Mgb      &      2.079 & 0.085& 0.034& 0.032\\
Fe5270   &    1.503 & 0.097 & 0.123 & 0.035\\
Fe5335     &   1.448 & 0.122 & 0.079 & 0.043\\

 \hline
  
\end{tabular}
\end{table}

\begin{table}
\caption{ Linear fit parameters (Y=a+bx) in the bulge region of NGC~2681}
\label{ajuste10}
\begin{tabular}{l l l l l}     
\hline \hline
Index & a  & err a & b  & err b \\ 
 HdA&5.205& 0.102&-0.390& 0.048\\
 HdF&  3.830  &0.062&-0.202&  0.030\\
 CN1& -0.082&  0.002 &0.003&  0.001\\
CN2&   -0.0439  &0.002  &0.003 &0.001\\
Ca4227& 0.617& 0.019& -0.006&0.010\\
 G4300& 1.922 & 0.065&  0.196  &0.034\\
 HgA&   2.535& 0.102&-0.446&0.051\\
 HgF&    3.289 & 0.045 &-0.243 & 0.024\\
 Fe4383&  2.736& 0.075&-0.009&  0.036\\
 Ca4455&  0.851 & 0.020 &-0.014 &0.010\\
Fe4531&   2.931& 0.041&-0.067&0.021\\
C4668&  4.825 & 0.0806&-0.214 &0.041\\
Hbeta&     4.230& 0.040&-0.190  &0.021\\
Fe5015&    5.504 &0.076& -0.211 & 0.038\\
Mg1&       0.034& 0.002& -0.001&  0.001\\
Mg2&       0.133 & 0.002& -0.001  &0.001\\
Mgb&       2.204& 0.022&-0.028&0.012\\
Fe5270&    2.386 & 0.039&-0.047&0.020\\
Fe5335&    2.294& 0.021&-0.101&0.012\\
 \hline
  
\end{tabular}
\end{table}

\begin{table}
\caption{ Linear fit parameters (Y=a+bx) in the bulge region of NGC~2859}
\label{ajuste11}
\begin{tabular}{l l l l l}     
\hline \hline
Index & a  & err a & b  & err b \\ 
HdA &-2.297& 0.105& 0.147& 0.042\\
HdF &0.214 & 0.073&  0.057 &0.030\\
CN1& 0.094&  0.002& -0.009 &0.001\\
CN2&  0.128 & 0.003 &-0.010& 0.001\\
Ca4227& 1.087 &0.027&-0.019&  0.011\\
G4300& 5.426  &0.090&-0.0588&  0.038\\
HgA&  -6.286& 0.127 & 0.163 &0.049\\
HgF& -1.579 &0.054 & 0.042&0.021\\
Fe4383& 5.583&  0.099& -0.178&0.040\\
Ca4455&  1.327& 0.034&-0.022 & 0.014\\
Fe4531&   3.450&0.078&-0.077&  0.032\\
C4668&   7.345 & 0.109 &-0.221  &0.043\\
Hbeta&     2.000& 0.047&-0.022 &0.019\\
Fe5015&   5.918 & 0.090& -0.159&0.035\\
Mg1&        0.121& 0.003& -0.004&0.001\\
Mg2&       0.274 & 0.002& -0.006&  0.001\\
Mgb&         4.252 & 0.039 &-0.052  &0.016\\
Fe5270&    3.070& 0.0411 &-0.032 &0.017\\
Fe5335&    2.916&0.037&-0.082 & 0.015\\
 \hline
  
\end{tabular}
\end{table}

\begin{table}
\caption{ Linear fit parameters (Y=a+bx) in the bulge region of NGC~2935}
\label{ajuste12}
\begin{tabular}{l l l l l}     
\hline \hline
Index & a  & err a & b  & err b \\ 
HdA&1.001 &0.228&  0.092& 0.080\\
 HdF&  1.411& 0.123&  0.104& 0.043\\
CN1&  0.045 & 0.005 &-0.004 & 0.002\\
 CN2&  0.069&  0.005& -0.002 & 0.002\\
Ca4227&0.705 & 0.053&-0.009& 0.018\\
G4300& 3.402&0.119&-0.048&  0.044\\
HgA& -2.848 &0.295  &0.207 & 0.104\\
HgF&  -0.128 &0.184&0.145& 0.065\\
Fe4383&  4.245&0.112&-0.143 & 0.041\\
Ca4455&  1.009 &0.064&-0.055 & 0.023\\
Fe4531&   2.725 &0.089 &-0.067&  0.034\\
C4668&  5.757  &0.172 &-0.248 &0.063\\
Hbeta&     2.196& 0.089& -0.017 & 0.032\\
Fe5015&   5.622 &0.210& -0.209&0.073\\
Mg1&      0.080& 0.003&  0.006& 0.001\\
Mg2&       0.199 & 0.004 &0.005&  0.001\\
Mgb&       2.807& 0.141&-0.051  &0.049\\
Fe5270&   2.632  &0.122&-0.052  &0.042\\
Fe5335&   2.243& 0.097 &-0.010&  0.034\\
 \hline
  
\end{tabular}
\end{table}

\begin{table}
\caption{ Linear fit parameters (Y=a+bx) in the bulge region of NGC~2950}
\label{ajuste13}
\begin{tabular}{l l l l l}     
\hline \hline
Index & a  & err a & b  & err b \\ 
HdA& -1.959& 0.074& 0.178 &0.024\\
HdF&    0.397 & 0.036&  0.052 & 0.012\\
CN1 &  0.112& 0.005&-0.015& 0.001\\
CN2    &0.149 & 0.005 &-0.016 & 0.001\\
Ca4227&  0.967&  0.029&-0.009& 0.010\\
G4300  &5.091 & 0.067 &-0.051 & 0.024\\
HgA    & -5.495 & 0.061  &0.126 & 0.020\\
HgF     & -1.193 & 0.027 &0.053 & 0.009\\
Fe4383 &   5.336& 0.061&-0.147 & 0.021\\
Ca4455   & 1.430& 0.038 &-0.061 & 0.013\\
Fe4531   & 3.559& 0.054 &-0.087& 0.019\\
C4668   & 9.607 & 0.175&-0.662&0.059\\
Hbeta    &  2.150& 0.031&-0.007 & 0.011\\
Fe5015   & 6.422 & 0.070 &-0.263&0.025\\
Mg1       & 0.121& 0.002&-0.005& 0.001\\
Mg2        & 0.272&  0.003 &-0.009 & 0.001\\
Mgb         &4.199 & 0.038&-0.122 &0.013\\
Fe5270    &3.271  &0.049& -0.109 &0.017\\
Fe5335   &3.055 &0.055&-0.103 & 0.019\\
\hline
  
\end{tabular}
\end{table}

\begin{table}
\caption{Linear fit parameters (Y=a+bx) in the bulge region of NGC~2962}
\label{ajuste14}
\begin{tabular}{l l l l l}     
\hline \hline
Index & a  & err a & b  & err b \\ 
HdA   &-3.122& 0.126& 0.405& 0.067\\
HdF  & -0.068& 0.076 & 0.111& 0.039\\
CN1  & 0.138& 0.005& -0.026 & 0.002\\
CN2   & 0.173 &0.006 &-0.028& 0.003\\
Ca4227& 1.308&0.056&-0.053& 0.030\\
G4300&   5.264&0.109&0.056& 0.057\\
HgA  & -6.730  &0.134 &0.176 & 0.068\\
HgF   &-1.920&0.093 & 0.093& 0.048\\
Fe4383& 5.969&0.1076 &-0.303& 0.057\\
Ca4455  &1.575  &0.058& -0.128& 0.030\\
Fe4531& 3.738& 0.102 &-0.186&  0.052\\
C4668 &8.918  &0.221 &-0.655 & 0.113\\
Hbeta& 1.615 &0.057 &0.072&  0.029\\
Fe5015 &6.271 &0.143&-0.321& 0.073 \\
Mg1     & 0.146&0.003& -0.009& 0.001\\
Mg2     & 0.314&0.003& -0.017&  0.001\\
Mgb     & 4.895 &0.055&-0.228& 0.029\\
Fe5270 & 3.373 &0.076&-0.128& 0.040\\
Fe5335&  3.258&  0.074&-0.127&  0.041\\
\hline
  
\end{tabular}
\end{table}

\begin{table}
\caption{ Linear fit parameters (Y=a+bx) in the bulge region of NGC~3081}
\label{ajuste15}
\begin{tabular}{l l l l l}     
\hline \hline
Index & a  & err a & b  & err b \\ 
HdA  &-4.618 &0.232& 0.994 & 0.103\\
HdF   & -2.286&0.164& 0.632 &0.072\\
CN1  & 0.214 &0.007 &-0.034&  0.003\\
CN2  & 0.214&0.007& -0.029&  0.003\\
Ca4227 &0.121&  0.140 & 0.190 & 0.065\\
G4300 &1.446 & 1.056& 0.922& 0.512\\
HgA    & -2.770& 1.726& -0.799 &0.765\\
HgF     & -3.906& 0.120& 0.893 &0.055\\
Fe4383 & 8.173 & 0.179& -1.012 & 0.100\\
Ca4455 & 0.525& 0.108& 0.150&  0.055\\
Fe4531 & 3.132& 0.375&-0.135& 0.184\\
C4668  &4.697& 0.301& 0.212& 0.148\\
Hbeta    &0.233& 0.181&0.308& 0.076\\
Fe5015  &0.233& 0.181&0.308& 0.076\\
Mg1      &0.293& 0.023&-0.044&0.009\\
Mg2      &0.357& 0.010& -0.028& 0.005\\
Mgb     &3.993&0.096 &-0.340 & 0.052\\
Fe5270 & 2.927& 0.102& -0.071&  0.051\\
Fe5335&  3.083 & 0.096&-0.229&0.049\\
\hline
  
\end{tabular}
\end{table}

\begin{table}
\caption{ Linear fit parameters (Y=a+bx) in the bulge region of NGC~4245}
\label{ajuste16}
\begin{tabular}{l l l l l}     
\hline \hline
Index & a  & err a & b  & err b \\ 
HdA& -3.068 & 0.184&0.928  &0.072\\
HdF&  -0.1189 & 0.079 &0.439&  0.031\\
CN1& 0.095& 0.006& -0.026&  0.002\\
CN2& 0.130 &0.005&-0.029&0.002\\
Ca4227& 1.369& 0.055& -0.103&0.022\\
G4300& 5.493 &0.149&-0.460&0.06\\
HgA& -6.673 &0.236  &0.969 &0.091\\
HgF& -1.642 &0.135 &0.489 & 0.052\\
Fe4383& 5.936&  0.153 &-0.393&  0.062\\
Ca4455& 1.498& 0.057&-0.083&  0.023\\
Fe4531& 3.608 & 0.112&-0.143 &0.046\\
C4668&   8.931 &0.217& -0.823 &0.089\\
Hbeta&    1.973& 0.072  &0.121&  0.029\\
Fe5015&  6.406& 0.152&-0.469&0.062\\
Mg1&    0.118&0.003&-0.011&  0.001\\
Mg2&   0.280& 0.004&-0.021& 0.001\\
Mgb&       4.444  &0.059& -0.302&0.024\\
Fe5270&   3.445& 0.074& -0.176& 0.031\\
Fe5335&  3.327 & 0.150 &-0.145&0.061\\
\hline
  
\end{tabular}
\end{table}
\clearpage

\begin{table}
\caption{ Linear fit parameters (Y=a+bx) in the bulge region of NGC~4314}
\label{ajuste17}
\begin{tabular}{l l l l l}     
\hline \hline
Index & a  & err a & b  & err b \\ 
HdA&-3.065&0.162&0.733 &0.042\\
HdF&  -0.078  &0.104&  0.310&0.027\\
CN1&  0.085 & 0.003& -0.020& 0.001\\
CN2& 0.118 &0.004&-0.021&0.001\\
Ca4227& 1.462&0.041&-0.125&  0.012\\
G4300& 5.706&0.194&-0.491&  0.057\\
HgA&  -7.428& 0.300 &1.027&0.081\\
HgF&   -2.046&0.165&0.506 &0.046\\
Fe4383& 6.378&0.162&-0.514 &0.050\\
Ca4455& 1.490&0.060& -0.105 &0.019\\
Fe4531& 3.651&0.130&-0.200 &0.040\\
C4668&  7.769&0.213&-0.593&0.067\\
Hbeta& 1.786&0.075&0.091&0.027\\
Fe5015& 6.143&  0.150&-0.367&0.048\\
Mg1&0.122 &0.002&-0.010&  0.001\\
Mg2&0.289 &0.004&-0.020&  0.001\\
Mgb&4.541&0.097&-0.265&0.031\\
Fe5270& 3.276&0.086&-0.133 &0.030\\
Fe5335& 3.134& 0.089&-0.171 &0.029\\
\hline

\end{tabular}
\end{table}
  
\begin{table}
\caption{ Linear fit parameters (Y=a+bx) in the bulge region of NGC~4394}
\label{ajuste18}
\begin{tabular}{l l l l l}     
\hline \hline
Index & a  & err a & b  & err b \\ 
HdA&  -1.694&  0.167&  0.985 &0.072\\
HdF&  0.580&0.116& 0.451& 0.050\\
CN1&  0.051&0.004& -0.027&  0.002\\
CN2&  0.084 & 0.004&-0.028  &0.002\\
Ca4227& 1.017& 0.036& -0.041&  0.016\\
G4300&   5.051&0.114&-0.440&0.054\\
HgA&    -5.368&0.230&1.102 &0.105\\
HgF;&    -1.051&0.109&  0.547&0.050\\
Fe4383&   5.381& 0.098&-0.482& 0.044\\
Ca4455&   1.404& 0.049&-0.150&  0.023\\
Fe4531&    3.557& 0.070&-0.228& 0.033\\
C4668&     7.653 & 0.142& -0.851&0.068\\
Hbeta&      2.214& 0.069  &0.161 &0.031\\
Fe5015&    5.530& 0.124 &-0.202&  0.058\\
Mg1&         0.090 &0.004 &-0.010& 0.001\\
Mg2&        0.238&  0.004& -0.017& 0.001\\
Mgb&         3.918& 0.077 &-0.276&  0.036\\
Fe5270&     3.216& 0.059&-0.210 & 0.028\\
Fe5335&      2.953 & 0.070& -0.200& 0.033\\
\hline

\end{tabular}
\end{table}

\begin{table}
\caption{ Linear fit parameters (Y=a+bx) in the bulge region of NGC~5101
\label{ajuste19}}
\begin{tabular}{l l l l l}     
\hline \hline
Index & a  & err a & b  & err b \\ 
HdA&  -3.675&  0.146& 0.173 &0.048\\
HdF&    -0.393 & 0.053  &0.059 & 0.017\\
CN1&   0.178 &0.003&-0.008& 0.001\\
CN2&    0.209& 0.004& -0.008& 0.001\\
Ca4227& 1.1481  &0.052& -0.013 & 0.018\\
G4300&  5.586& 0.052 &-0.039& 0.018\\
HgA&    -7.671&  0.075&0.137 & 0.025\\
HgF&      -2.466&  0.063& 0.059 &0.021\\
Fe4383&  6.326& 0.120& -0.170& 0.042\\
Ca4455&  1.394&  0.046& -0.021& 0.016\\
Fe4531&   3.446 & 0.087 &-0.031& 0.030\\
C4668&    8.434&  0.098&-0.202&  0.034\\
Hbeta&     1.891  &0.053& -0.107& 0.019\\
Fe5015&   6.762&  0.146 &-0.182& 0.051\\
Mg1&       0.139 & 0.001& 0.005& 0.001\\
Mg2&        0.308&  0.002&  0.003&0.001\\
Mgb&        4.168 & 0.087 &-0.041  &0.030\\
Fe5270&    3.331&  0.059& -0.038&  0.020\\
Fe5335&    3.069& 0.062&-0.075&  0.022\\
\hline
  
\end{tabular}
\end{table}

\end{document}